\documentclass[11pt]{article}
\usepackage[utf8]{inputenc}
\usepackage{amsthm,amsmath,amssymb,graphicx,gensymb} 
\RequirePackage[numbers]{natbib}
\RequirePackage[colorlinks,citecolor=blue,urlcolor=blue]{hyperref}
\usepackage[dvipsnames]{xcolor}
\usepackage{color}
\usepackage{bm,bbm}
\usepackage[margin=2cm]{geometry}

\DeclareMathOperator*{\argmin}{arg\,min}

\title{\hrule\vspace{10pt} Estimating the limiting shape of bivariate scaled sample clouds: with additional benefits of self-consistent inference for existing extremal dependence properties}
\author{Emma S.\ Simpson$^{1\footnote{Corresponding author. Email address: emma.simpson@ucl.ac.uk}}$ ~and Jonathan A.\ Tawn$^2$\\ 
\normalsize{$^1$Department of Statistical Science, University College London, WC1E 6BT, U.K.}\\
\normalsize{$^2$School of Mathematical Sciences, Lancaster University, LA1 4YF, U.K.}}
\date{\today\vspace{10pt}\hrule\vspace{-0.5cm}}

\begin{document}
\maketitle

\begin{abstract}
The key to successful statistical analysis of bivariate extreme events lies in flexible modelling of the tail dependence relationship between the two variables. In the extreme value theory literature, various techniques are available to model separate aspects of tail dependence, based on different asymptotic limits. Results from Balkema and Nolde \cite{Balkema2010} and Nolde \cite{Nolde2014} highlight the importance of studying the limiting shape of an appropriately-scaled sample cloud when characterising the whole joint tail. We now develop the first statistical inference for this limit set, which has considerable practical importance for a unified inference framework across different aspects of the joint tail. Moreover, Nolde and Wadsworth \cite{Nolde2021} link this limit set to various existing extremal dependence frameworks. Hence, a by-product of our new limit set inference is the first set of self-consistent estimators for several extremal dependence measures, avoiding the current possibility of contradictory conclusions.  In simulations, our limit set estimator is successful across a range of distributions, and the corresponding extremal dependence estimators provide a major joint improvement and small marginal improvements over existing techniques. We consider an application to sea wave heights, where our estimates successfully capture the expected weakening extremal dependence as the distance between locations increases.
\end{abstract}
\noindent{\bf Keywords:} bivariate extremes; coefficient of asymptotic independence; conditional extremes; extremal dependence structure; gauge function; sample cloud\\

\noindent{\bf Acknowledgements} This publication is based upon work supported by the King Abdullah University of Science and Technology (KAUST) Office of Sponsored Research (OSR) under Award No.\ OSR-2017-CRG6-3434.02. We thank Philip Jonathan and Jennifer Wadsworth for providing access to the data studied in Section~\ref{subsec:waves}.

\newpage\section{Introduction}\label{sec:intro}
Multivariate extreme value problems are important across a range of subject domains, such as sea level \cite{Coles1994}, air pollution \cite{Heffernan2004}, rainfall \cite{Davison2012} and river flow \cite{Engelke2020}. The typical formulation is to have $n$ independent and identically distributed observations, $(\bm{x}_1,\ldots ,\bm{x}_n)$, from a $d$-dimensional random vector $\bm{X}\in\mathbbm{R}^d$ with unknown joint distribution $F_{\bm{X}}$. Generally, the aim is to estimate $F_{\bm{X}}(\bm{x})$ and $\Pr(\bm{X}\in A)$ for some $A\subset \mathbbm{R}^d$, where $\bm{x}$ and all the elements in $A$ are large in at least one component of $\bm{X}$. Our focus here is slightly different; we instead aim to estimate the boundary of a set that in the limit as $n\rightarrow \infty$ contains all sample values from $\bm{X}$ after some appropriate scaling.
The purpose of this is two-fold: (i) to identify potentially `risky' combinations of variables in finite samples and (ii) to provide improved inference for existing extremal dependence measures. 

The usual approach to inference in multivariate extremes is to estimate the marginal distributions and dependence structure (copula) with a focus on their behaviour in the tail region. Univariate extreme value methods are well established \cite{Coles2001, Davison1990}, with dependence modelling being the key challenge. A standard framework is to transform the margins of $\bm{X}$ to a common distribution based on univariate extreme value models fitted to the individual components. In extreme value analysis, common marginal selections include Fr\'{e}chet, Gumbel, Exponential and Laplace distributions, with choices made based on theoretical results and/or to highlight particular tail dependence features. Here, our focus is on Exponential marginals, although we discuss possible extensions to other choices in Section~\ref{sec:discussion}, and we take $d=2$ for ease of exposition. That is, we consider variables $(X_1,X_2)$ with $\Pr(X_j< x)=1-\exp(-x)$, for $x\geq 0$ and $j=1,2$.

Even with common margins, multivariate extremes are often difficult to model due to the lack of a natural ordering \citep{Barnett1976}. As such, there have been several different asymptotically-motivated modelling approaches and assumptions proposed in the literature, e.g., multivariate regular variation \cite{Resnick1987}, hidden regular variation \cite{Ledford1997, Resnick2002}, conditional extremes \cite{Heffernan2004, Heffernan2007}, powered joint tails \cite{Wadsworth2013,deValk2016}, and mixture structures \cite{Goix2017,Simpson2020}. The asymptotic arguments differ in each of these cases, covering situations where the growth rates of the margins are the same or different across all variables, and where they can vary with the direction of the extreme region of interest. 

Results from \cite{Balkema2010} on the limiting shape, and associated boundary $G$, of an appropriately-scaled sample cloud offer a new, more unified, asymptotic approach. Here, the focus is on the scaled sample cloud, corresponding to $n$ independent samples from the joint distribution of $(X_1,X_2)$, denoted by 
\begin{equation} 
C_n=\{(X_{1,i}/\log n,X_{2,i}/\log n); i=1,\dots,n\}.
\label{eqn:scaledsamplecloud}
\end{equation}
We note that the $\log$-scaling is appropriate here due to $\bm{X}$ having Exponential margins; we comment on this further in Section~\ref{subsec:theory}. As $n\rightarrow \infty$, the sample $C_n$ converges onto a compact limit set with upper boundary  
\begin{equation}
    G=\{(x_1,x_2)\in[0,1]^2:g(x_1,x_2)=1\},
    \label{eqn:G}
\end{equation}
for a function $g$ termed the gauge function, which will also be defined in Section~\ref{subsec:theory}. Given that $\bm{X}$ has Exponential margins, the shape of $G$ is fully determined by the extremal dependence of the variables. Here, $G$ provides a characterisation of the full joint tail, irrespective of the nature of extremal dependence, whereas  the previously mentioned methods offer a separate treatment for different forms of extremal dependence. Balkema and Nolde \cite{Balkema2020} discuss, at a purely probabilistic level, how this limit set can be used for risk evaluation and tail inference for linear combinations of $\bm{X}$.

Although $G$ is a limit set boundary, we propose that it can also be used to provide valuable information about the likely boundary set for finite samples.  
Specifically, the boundary set
\begin{align}
    G_m:=\{(x_1, x_2)\in \mathbbm{R}^2_+:g(x_1/\log m,x_2/\log m)=1\}
\label{eqn:finiteSampleG}
\end{align}
for $m\in \mathbbm{N}$, is not exceeded by any pair in the sample $\{(X_{1,i},X_{2,i}): i=1,\ldots ,m\}$ with probability tending to $1$ as $m\rightarrow \infty$, and 
with no other curve satisfying this property if it has on it any $(x_1,x_2)$ that is below $G_m$. Hence, for large $m$, this boundary set can be used to identify the most risky combinations of variables for a sample size of $m$, and when $m>n$, extrapolation for the purposes of risk assessment is possible.

The statistical exploitation of this limit theory through inference for $G$, which is the aim of this paper, has not been explored but has two key benefits. First, there is considerable practical advantage in the estimator $\hat{G}_m$ of $G_m$, obtained via estimation of~\eqref{eqn:G}, using the sample of size $n$ for extrapolation purposes.  Secondly, as \cite{Nolde2021} have shown, different parts of the boundary set $G$ provide valuable geometric interpretations for a number of existing extremal dependence measures, discussed in Section~\ref{subsec:BEVfeatures}, which play key roles in inference for the multivariate extremes methods listed above. An additional benefit of estimating $G$ is that it provides new estimators of these dependence measures through exploitation of these theoretical results. It is therefore of interest to compare how each of the resulting estimators, based on a unifying estimate $\hat{G}$ of $G$, performs individually, relative to existing techniques. Critically, though, we see most value in using $\hat{G}$ to ensure self-consistency over the estimated dependence measures. The previous estimators do not guarantee such self-consistent information, since they are estimated separately using different asymptotic theories.

To illustrate the issue of having a lack of self-consistency in the estimation of extremal dependence features, we consider the most common starting point for a bivariate extreme value analysis: identifying whether $(X_1,X_2)$ are \emph{asymptotically dependent} or \emph{asymptotically independent} \cite{Coles1999}. For our Exponentially distributed variables, this can be formalised through the coefficient of asymptotic dependence $\chi$, with
\begin{equation}
\chi= \lim_{x\rightarrow \infty}\Pr\left(X_2>x\mid X_1>x\right)= \lim_{x\rightarrow \infty}\frac{\Pr\left(X_1>x, X_2>x\right)}{\exp(-x)}.
\label{eqn:chi}
\end{equation}
A value of $\chi>0$ $(\chi=0)$ corresponds to asymptotic dependence 
(asymptotic independence) between $X_1$ and $X_2$, respectively. From limit~\eqref{eqn:chi}, it can be seen that asymptotic dependence occurs when the most extreme values can happen simultaneously in the two variables, and asymptotic independence is when such joint extremes are not possible. 
Since many asymptotic results for bivariate extremes, and their associated statistical models, are only suitable in one of these situations, distinguishing between them can play a crucial role in selecting a modelling approach. With the existing estimators it is possible that, for example, the coefficient of asymptotic independence of \cite{Ledford1996} indicates asymptotic dependence ($\chi>0$) while the normalising functions in the conditional extremes approach of \cite{Heffernan2004} point towards asymptotic independence ($\chi=0$). With our proposed estimators, unhelpful contradictions such as this are avoided; we return to this point in Section~\ref{subsec:properties}.

Finally, of course, there is much more to understanding bivariate extremes than simply distinguishing between $\chi>0$ and $\chi=0$, with inference for $G$ helping here as well. For example, irrespective of whether the variables are asymptotically dependent or asymptotically independent, it is possible that one variable could take its largest values while the other is smaller order; a scenario of importance when studying the nature of multivariate extreme events \cite{Goix2017, Simpson2020}. In such cases, models that are able to capture a mixture structure in the extremal dependence features are required, with \cite{Tendijck2021} giving a first approach. Inference for $G$ provides key diagnostics for identifying this possibility in a way that is self-consistent with respect to $\chi$ and other extremal dependence features.

We continue the paper by providing further detail and examples on the asymptotic scaled sample cloud in Section~\ref{subsec:limitSet}, while definitions of various extremal dependence measures and their links to $G$ are covered in Section~\ref{sec:theory}. We introduce our procedure for estimating $G$ in Section~\ref{sec:method} and explain how this can be exploited to obtain self-consistent estimators of the considered extremal dependence properties in Section~\ref{sec:parameterEsts}. The performance of our estimation procedure is demonstrated through a simulation study and application to sea wave heights in Section~\ref{sec:results}. We conclude with a discussion in Section~\ref{sec:discussion}.

\section{The limiting shape of a scaled sample cloud}\label{subsec:limitSet}
\subsection{General theory}
\label{subsec:theory}
A sample cloud corresponds to $n$ independent replicates from the joint distribution of $(X_1,X_2)$; we denote this by $C^*_n=\{(X_{1,i},X_{2,i}); i=1,\dots,n\}$. The limiting convex hull of $C^*_n$, as $n\rightarrow\infty$, has been previously studied; see for example \cite{Eddy1981, Brozius1987, Davis1987}. Following \cite{Nolde2021}, our focus is on a scaled version of $C^*_n$, which we denote $C_n$, as defined in expression~\eqref{eqn:scaledsamplecloud}. We assume that we have standard Exponential marginal distributions, and that the joint density $f(x_1,x_2)$ exists and is non-zero everywhere in $\mathbbm{R}^2_+$. The $\log n$ scaling is chosen due to the marginal distributions of $X_1$ and $X_2$, as it ensures that $\max_{i=1,\dots,n}\left(X_{j,i}/\log n\right)\xrightarrow{p}1$, as $n\rightarrow\infty$, for $j=1,2$; \cite{Nolde2014} shows that other light-tailed margins could also be used here and would result in different scaling functions being required, as well as affecting the shape of $G$. 

Interest lies in studying the asymptotic shape of $C_n$, as $n\rightarrow\infty$. A useful tool in this task is the gauge function $g(x_1,x_2)$, defined via the relationship
\begin{equation}
-\log f(tx_1,tx_2) \sim tg(x_1,x_2),\qquad t\rightarrow\infty,\qquad x_1,x_2\geq0;
\label{eqn:gauge}
\end{equation}
see \cite{Nolde2014,Nolde2021}. As $n\rightarrow\infty$, $C_n$ converges onto the compact limit set 
\[
    G_S=\{(x_1,x_2)\subseteq[0,1]^2 :g(x_1,x_2)\leq 1\}.
\] 
The set $G_S$ is star-shaped, meaning that for any $(x_1,x_2)\in G_S$, $(vx_1,vx_2)\in G_S$ for all $v\in(0,1)$. In the following, we assume that the limit set $G_S$ exists. 

The boundary of $G_S$, defined by the set $G$ in~\eqref{eqn:G}, is of particular interest. Some important properties of $G$ are as follows: (i) due to the log-scaling, $G$ must touch the lines $x_1=1$ and $x_2=1$ at least once; (ii) if $f$ is non-zero everywhere, $G$ must be a continuous function; and (iii) by limit~\eqref{eqn:gauge}, the gauge function must be homogeneous of order 1, so that, for example, taking $r=x_1+x_2$ and $w=x_1/r$, with $(x_1, x_2) \in G$, we have $g(rw,r(1-w))=rg(w,1-w)$. These properties will be crucial in the development of our inferential approach, presented in Section~\ref{sec:method}.

\subsection{Theoretical examples}
\label{subsec:Examples}
\begin{figure}[t]
    \centering
    \includegraphics[width=11cm]{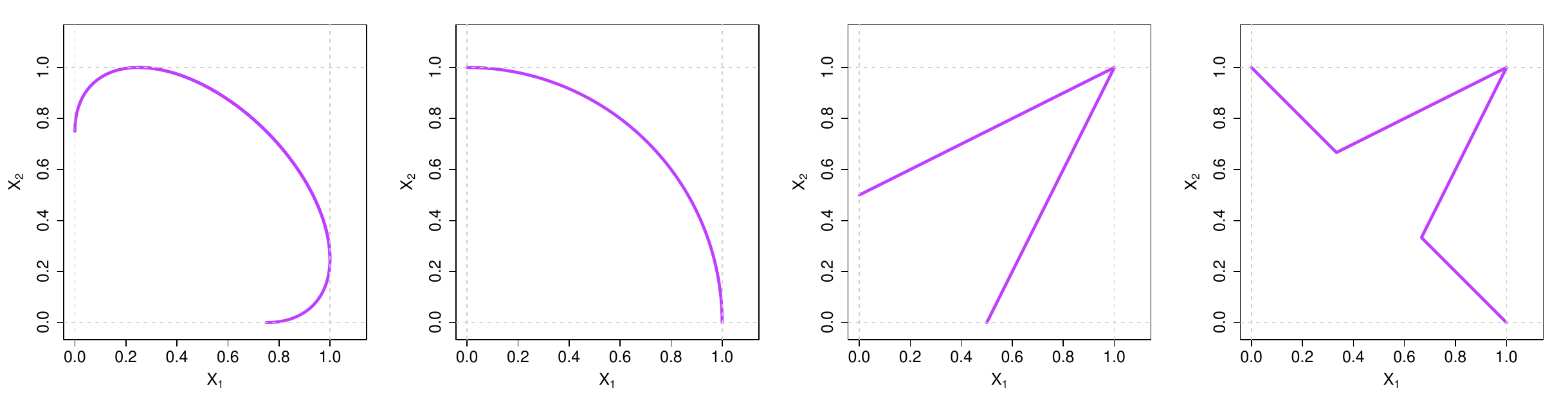}
    \caption{The boundary set $G$ for Gaussian, inverted logistic, logistic and asymmetric logistic (left to right) dependence models with Exponential margins. The Gaussian correlation and the (inverted/asymmetric) logistic dependence parameters are set to $\rho=\gamma=0.5$.}
    \label{fig:gaugeEx}
\end{figure}

The shapes of the limit set $G_S$ and its boundary $G$ are determined by the dependence between the variables. We illustrate these sets through four well-known copula families (see Figure~\ref{fig:gaugeEx}) which show that the general properties of $G_S$ and $G$ stated in Section~\ref{subsec:theory} hold, as well as providing an indication of the variety of different shapes that $G$ can take. We use these distributions as examples throughout the paper. Given their known extremal dependence properties, these four examples provide insight into how $G$ captures these features, with further details on this link presented in Section~\ref{sec:NWtheory}.

The first example is a bivariate Gaussian copula with correlation parameter $\rho\in[0,1]$. Nolde \cite{Nolde2014} shows that the gauge function in this case has the form
\[
g(x_1,x_2) = (1-\rho^2)^{-1}\left(x_1+x_2-2\rho x_1^{1/2}x_2^{1/2}\right), \qquad x_1,x_2\geq 0.
\]
In the case of independence between the variables, i.e., when $\rho=0$, the gauge function is simply $g(x_1,x_2)=x_1+x_2$. For details on the calculation of the remaining three gauge functions discussed below, see \cite{Simpson2021}. First, consider an inverted bivariate extreme value copula with logistic model and dependence parameter $\gamma\in(0,1]$ \citep[see][]{Ledford1997}, with $\gamma=1$ corresponding to complete independence, which has a gauge function of the form
\[
g(x_1,x_2) = \left(x_1^{1/\gamma}+x_2^{1/\gamma}\right)^\gamma, \qquad x_1,x_2\geq 0,
\]
and a bivariate extreme value copula with logistic model \citep{Gumbel1960}, again with dependence parameter $\gamma\in(0,1]$, having the gauge function
\begin{equation*}
g(x_1,x_2) = 
    \begin{cases}
        \frac{1}{\gamma}\max(x_1,x_2)-\left(\frac{1}{\gamma}-1\right)\min(x_1,x_2), \qquad &x_1,x_2\geq 0, \gamma<1,\\
        x_1+x_2, \qquad &x_1,x_2\geq 0, \gamma=1.
    \end{cases}
\end{equation*}
There is a discontinuity here between $\gamma=1$ and $\gamma\rightarrow 1$, which is linked to findings about the logistic model in \cite{Ledford1996}; since $\gamma=1$ corresponds to independence, which is already covered by the Gaussian copula with $\rho=0$, we restrict to the $\gamma<1$ case in the remainder of the paper. 

The first two examples in Figure~\ref{fig:gaugeEx} are widely known to correspond to asymptotically independent models, while the third exhibits asymptotic dependence if $\gamma<1$ \citep{Coles1999}. An important link between the bivariate gauge function and extremal dependence comes from the value of $g(1,1)$ \citep{Nolde2014}. If $g(1,1)>1$, this corresponds to the case of asymptotic independence with $\chi=0$, while under asymptotic dependence, with $\chi>0$, we have $g(1,1)=1$. This is clearly satisfied by the three examples above. However, it is also possible to have $g(1,1)=1$ and $\chi=0$, so care with interpretation is needed.

Finally, we consider a bivariate extreme value copula with asymmetric logistic model \citep{Tawn1988}, where a mixture of extremal dependence features is possible, arising from logistic and independence components. That is, both variables can be simultaneously extreme, putting this model in the asymptotic dependence class, but each of the variables can also be individually large while the other is of smaller order. We provide details on the form of this dependence model in the supplementary material \cite[Section I]{SupplementaryMaterial}. Following the discussion of discontinuity in the logistic model above, we again restrict the dependence parameter to $\gamma\in(0,1)$. Moreover, setting the additional model parameters to $\theta_1=\theta_2=0$ yields a logistic model, while setting $\theta_1=\theta_2=1$ recovers the setting of independence; we therefore focus on $\theta_1,\theta_2\in(0,1)$ in the following. Figure~\ref{fig:gaugeEx} (right panel) also shows the gauge function of this copula, which is of the form
\begin{equation}
g(x_1,x_2) = \min\left\{(x_1+x_2); \frac{1}{\gamma}\max(x_1,x_2)-\left(\frac{1}{\gamma}-1\right)\min(x_1,x_2)\right\}, ~ x_1,x_2\geq 0.
\label{eqn:alogGauge}
\end{equation}
Here, we once again have $\chi>0$ and it is clear that $g(1,1)=1$. Interestingly, the gauge function does not depend on the parameters $\theta_1$ and $\theta_2$ when $\theta_1,\theta_2\in(0,1)$.

\section{Links between the sample cloud boundary and extremal dependence measures}\label{sec:theory}

\subsection{Bivariate extremal dependence features}\label{subsec:BEVfeatures}
We now introduce the tail dependence features that we aim to estimate in a self-consistent way by ultimately exploiting properties of $G$. In Section~\ref{sec:NWtheory}, we explain how each of these dependence features is linked to the set $G$, using results from \cite{Nolde2021}.
We begin with the coefficient of asymptotic independence, $\eta$, of \cite{Ledford1996}, or equivalently $\bar\chi=2\eta-1$ \citep{Coles1999}. Recall that we assume the variables $X_1$ and $X_2$ have standard Exponential marginal distributions. Then, $\eta\in(0,1]$ is defined by considering the behaviour of the joint survivor function
\begin{equation*}
    \Pr(X_1>x,X_2>x)\sim \mathcal{L}(e^x)e^{-x/\eta},
\end{equation*}
as $x\rightarrow\infty$, where the function $\mathcal{L}$ is slowly varying at infinity. Considering the definition of $\chi$ in~\eqref{eqn:chi}, when $\eta=1$, this yields $\chi=\lim_{x\rightarrow\infty}\mathcal{L}(e^x)$, whereas if $\eta<1$, we obtain $\chi=\lim_{x\rightarrow\infty}\mathcal{L}(e^x)e^{-(1/\eta -1)x}=0$. Hence, for $\eta=1$ and $\mathcal{L}(x)\not\rightarrow 0$ as $x\rightarrow\infty$, we have $\chi>0$, and therefore asymptotic dependence. If $\eta<1$, or $\eta=1$ and $\mathcal{L}(x)\rightarrow 0$  as $x\rightarrow\infty$, we have $\chi=0$ and the variables are asymptotically independent. Estimation of the coefficient $\eta$ can contribute towards the classification of tail dependence behaviour in practice, and in turn enable the selection of an appropriate model for the joint extremes. 

Wadsworth and Tawn \cite{Wadsworth2013} extend the approach of Ledford and Tawn \cite{Ledford1996} by allowing for different scalings. They consider limiting probabilities of the form
\begin{equation*}
    \Pr\left\{X_1>\omega x,X_2>(1-\omega)x\right\}\sim \mathcal{L}_\omega(e^x)e^{-x\lambda(\omega)},
\end{equation*}
as $x\rightarrow\infty$, for $\omega\in[0,1]$, $\lambda(\omega)\in(0,1]$ and some $\mathcal{L}_\omega$ that is slowly varying at infinity. The case where $\omega=1/2$ is linked to the coefficient $\eta$ by the relation $\eta^{-1}=2\lambda(1/2)$. Therefore, under asymptotic dependence $\lambda(1/2)=1/2$, and more generally, for $\omega\in[0,1]$, we have $\lambda(\omega)=\max(\omega,1-\omega)$ in this case. Under complete independence, $\lambda(\omega)=1$ for all $\omega\in[0,1]$.

Motivated by the possibility of mixture structures in the extremal dependence features, Simpson et al.\ \cite{Simpson2020} introduced a further set of indices related to $\eta$. They consider a separate measure for each subset of variables, which describes whether they can be simultaneously large while the other variables are of smaller order. In the bivariate case, there are two measures of interest, denoted by $\tau_1(\delta),\tau_2(\delta)\in(0,1]$, for $\delta\in[0,1]$. The measure $\tau_1(\delta)$ is based on a hidden regular variation assumption, see \cite{Simpson2020} and \cite{Nolde2021}, and in Exponential margins is defined by the relation
\begin{equation}
\Pr(X_1>x,X_2\leq \delta x) \sim \mathcal{L}_\delta(e^x)e^{-x/\tau_1(\delta)},
\label{eqn:tauDef}
\end{equation}
as $x\rightarrow\infty$, for $\delta\in[0,1]$ and some slowly varying function $\mathcal{L}_\delta$. The function $\tau_1(\delta)$ is monotonically increasing with $\delta$, and has $\tau_1(1)=1$. If there exists any $\delta^*<1$ such that $\tau_1(\delta^*)=1$, the variable $X_1$ can take its largest values while $X_2$ is of smaller order, otherwise $X_1$ can only take its largest values when $X_2$ is also large. The measure $\tau_2(\delta)$ can be defined analogously through the limiting behaviour of $\Pr(X_1\leq \delta x,X_2>x)$, as $x\rightarrow\infty$. 

Finally, we consider the conditional extremes modelling approach of Heffernan and Tawn \cite{Heffernan2004}. Conditional extremes models capture both asymptotic dependence and asymptotic independence, so are widely applicable. The conditional extremes framework requires that the marginal distributions have Exponential upper tails, which is clearly satisfied when $(X_1,X_2)$ have standard Exponential margins. Selecting $X_1$ as the conditioning variable, we assume there exist functions $a_1(\cdot)$ and $b_1(\cdot)>0$ such that
\begin{equation}
    \left\{\frac{X_2-a_1(X_1)}{b_1(X_1)},X_1-u\right\}\bigg\vert X_1>u\rightarrow (Z,E),\qquad \text{as }u\rightarrow\infty,
    \label{eqn:HandT}
\end{equation}
where $E\sim Exp(1)$ is independent of $Z$, and $Z$ represents some non-degenerate residual distribution that places no mass on $\{+\infty\}$ \citep{Heffernan2004,Keef.al.2013}. Under this conditioning, a suitably normalised version of $X_2$ and exceedances of $X_1$ above the threshold $u$ become independent as $u\rightarrow\infty$. Heffernan and Tawn \cite{Heffernan2004} propose setting $a_1(x)=\alpha_1 x$ and $b_1(x)=x^{\beta_1}$, for $\alpha_1\in[0,1]$ and $\beta_1\in[0,1)$, and demonstrate that this is a reasonable choice for a range of non-negatively dependent distributions. The case where $\alpha_1=1$ and $\beta_1=0$ corresponds to asymptotic dependence, while $\alpha_1<1$ corresponds to asymptotic independence, with complete independence achieved when $\alpha_1=0$, $\beta_1=0$ and $Z\sim Exp(1)$. An analogous result holds if $X_2$ is selected as the conditioning variable, with normalising functions $a_2(x)=\alpha_2 x$ and $\beta_2(x)=x^{\beta_2}$ for $\alpha_2\in[0,1]$ and $\beta_2\in[0,1)$. 

\begin{table}[t]
\centering
\resizebox{12cm}{!}{
\begin{tabular}{|l|c|c|c|c|c|} \hline
      Dependence model & $\eta$ & $\lambda(\omega)$ &  $\tau_1(\delta)=\tau_2(\delta)$  & $\alpha_1=\alpha_2$ & $\beta_1=\beta_2$ \\ \hline
      Gaussian & $(1+\rho)/2$  & $\begin{cases}
                    \frac{1-2\rho\left\{\omega(1-\omega)\right\}^{1/2}}{1-\rho^2}, \text{ if } t_\omega\geq\rho^2\\ 
                    \max(\omega,1-\omega), ~\text{ if }t_\omega<\rho^2\\
                    \end{cases}$& 
                    $\begin{cases}
                    1, \text{ if } \delta\geq\rho^2\\ 
                    \frac{1 -\rho^2}{1+\delta-2\rho\delta^{1/2}}, \text{ if }\delta<\rho^2\\
                    \end{cases}$ & $\rho^2$ & $1/2$ \\
      Inverted logistic & $2^{-\gamma}$ & $\left\{\omega^{1/\gamma}+(1-\omega)^{1/\gamma}\right\}^\gamma$ & 1 & 0 & $1-\gamma$  \\
      Logistic & 1 & $\max(\omega,1-\omega)$ & $\gamma/(1+\gamma\delta-\delta)$ & 1 & 0  \\
      Asymmetric logistic & 1 & $\max(\omega,1-\omega)$ & 1 & 1 & 0 \\ \hline
    \end{tabular}}    
      \caption{Values of $\{\eta,\lambda(\omega),\tau_1(\delta),\tau_2(\delta),\alpha_1,\alpha_2,\beta_1,\beta_2\}$ for a Gaussian model with correlation parameter $\rho\in[0,1]$ and (inverted) logistic models with dependence parameter $\gamma\in(0,1)$. Here, $t_\omega=\min(\omega,1-\omega)/\max(\omega,1-\omega)$.}
      \label{tab:parameterValues}
\end{table}

The values of $\eta$, $\lambda(\omega)$, $\tau_1(\delta)$, $\tau_2(\delta)$, $\alpha_1$, $\alpha_2$, $\beta_1$ and $\beta_2$ for the copulas considered in Figure~\ref{fig:gaugeEx} are given in Table~\ref{tab:parameterValues}. For the Gaussian distribution, the $\tau_i(\delta)$ $(i=1,2)$ result for $\delta<\rho^2$ was derived by \cite{Nolde2021}. For each of the first three copula models, the exchangeability of the variables means that $\tau_1(\delta)=\tau_2(\delta)$, $\alpha_1=\alpha_2$ and $\beta_1=\beta_2$. Table~\ref{tab:parameterValues} also shows this property holds in the asymmetric logistic case when the variables are not exchangeable, i.e., when $\theta_1\neq\theta_2$. 
More generally, in Section~\ref{sec:NWtheory}, we will see that these equivalences also arise when $G$ is symmetric about the line $x_1=x_2$, even if the distribution itself is not exchangeable.

\subsection{Boundary set interpretation of bivariate extremal dependence features}\label{sec:NWtheory}
This section provides a summary of key information from the important theoretical work of Nolde and Wadsworth \cite{Nolde2021}, which established links between $G$ and the extremal dependence features from Section~\ref{subsec:BEVfeatures}. Balkema and Nolde \cite{Balkema2010} were the first to provide results linking the asymptotic shape of a scaled sample cloud to asymptotic independence. Nolde \cite{Nolde2014} extended this work to show that the coefficient of asymptotic independence, $\eta$, is linked to the set $G$ defined in~\eqref{eqn:G} via the relationship
\begin{equation}
\eta = \min\left\{s\in(0,1]: [s,\infty]^2\cap G=\emptyset\right\}.
\label{eqn:Neta} 
\end{equation}
We can think of this as moving the set $[1,\infty]^2$ along the line $x_2=x_1$ towards the origin, until it intersects the set $G$ (although this intersection does not have to occur on the diagonal). A pictorial demonstration of this result, and those that follow in this section, is provided in the supplementary material \cite[Section A]{SupplementaryMaterial}. A higher dimensional version of~\eqref{eqn:Neta} was previously used in \cite{Simpson2021} to study the extremal dependence structure of vine copulas.

To calculate the value of $\lambda(\omega)$, \cite{Nolde2021} consider sets of the form 
\[
S_\omega = \left\{(x_1,x_2): x_1 \ge \omega/\max(\omega,1-\omega), x_2\ge (1-\omega)/\max(\omega,1-\omega)\right\},
\]
for any fixed $\omega\in[0,1]$.
They prove that
\begin{equation}
    \lambda(\omega)=\frac{\max(\omega,1-\omega)}{s_\omega},\qquad \text{with } s_\omega = \min\left\{s\in[0,1]:sS_\omega\cap G =\emptyset\right\}.
\label{eqn:NWlambda}
\end{equation}
The set $S_{1/2}$ is equivalent to $[1,\infty]^2$, so that the value of $s_{1/2}$ is $\eta$, as we expect since $\lambda(1/2)=1/(2\eta)$, as discussed in Section~\ref{subsec:BEVfeatures}. Nolde and Wadsworth \cite{Nolde2021} link the set $G$ to $\tau_1(\delta)$ and $\tau_2(\delta)$ by considering sets
\begin{equation*}
S_{1,\delta} = \left\{(x_1,x_2): x_1\in(1,\infty], x_2\in [0,\delta]\right\}, 
\end{equation*}
\begin{equation*}
S_{2,\delta} = \left\{(x_1,x_2): x_1\in[0,\delta], x_2\in (1,\infty]\right\},
\end{equation*}
for $\delta\in[0,1]$, and showing that, for $i=1,2$,
\begin{equation}
\tau_i(\delta) = \min\left\{s\in(0,1]:sS_{i,\delta}\cap G =\emptyset\right\}.
\label{eqn:NWtau}
\end{equation}
Finally, in the conditional extremes framework, \cite{Nolde2021} show that 
\begin{equation}
    \alpha_1 = \max\left\{\tilde\alpha_1\in[0,1] : g(1,\tilde\alpha_1)=1\right\}.
\label{eqn:NWalpha}
\end{equation}
It is possible to have more than one such $\tilde\alpha_1$, but the largest value is needed to avoid $Z$ in~\eqref{eqn:HandT} placing mass on $\{+\infty\}$. This means we should find the largest value of $x_2$ where $G$ intersects the line $x_1=1$. Considering an analogous definition of $\alpha_2$, alongside result~\eqref{eqn:Neta}, it is clear that $\eta\geq\max(\alpha_1,\alpha_2)$. If there are $\nu$ separate values of $\tilde\alpha_1$ in~\eqref{eqn:NWalpha}, this indicates a $\nu$-component mixture structure in the extremal dependence conditioning on $X_1$ large, following the representation of \cite{Tendijck2021}. Moreover, the smallest value of $\delta$ such that $\tau_1(\delta)=1$ is given by $\delta^*=\min\left\{\tilde\alpha_1\in[0,1] : g(1,\tilde\alpha_1)=1\right\}$. If $\nu=1$ with only one value of $\tilde\alpha_1$ satisfying $g(1,\tilde\alpha_1)=1$, we have $\delta^*=\alpha_1$; this is the case for the first three examples in Figure~\ref{fig:gaugeEx}. The asymmetric logistic example in Figure~\ref{fig:gaugeEx} has $\nu=2$ and $\tilde\alpha_1\in\{0,1\}$, yielding $\alpha_1=1$ and $\delta^*=0$. Overall, the set of measures $\left\{\eta, \max_{\delta<1}\tau_1(\delta), \max_{\delta<1}\tau_2(\delta)\right\}$ can describe whether the variables $(X_1,X_2)$ can be simultaneously large, whether one variable is large while the other is of smaller order, or whether we have a combination of these cases, but considering the $\tilde\alpha_1$ values from~\eqref{eqn:NWalpha} can provide additional insight.

Nolde and Wadsworth \cite{Nolde2021} also present results on the parameter $\beta_1$ in the conditional extremes model. They show that
\begin{equation}
    g(1,\alpha_1 + u) = 1 + O\left\{u^{1/(1-\beta)}\right\}, \qquad \text{as }u\rightarrow 0.
\label{eqn:NWbeta}
\end{equation}
The results for $\alpha_2$ and $\beta_2$ are analogous to those presented here.

As well as considering how the extremal dependence features described above can be obtained from the set $G$, we examine how much information these features can give us about $G$. The indices $\lambda(\omega), \tau_1(\delta)$ and $\tau_2(\delta)$ ($\omega,\delta\in[0,1]$) are the most interesting from this perspective, as collectively they can sometimes fully describe $G$. In the logistic case, only $\tau_1(\delta)$ and $\tau_2(\delta)$ are needed; the opposite is true for the inverted logistic case; whereas for the Gaussian case, they are all required. However, for the asymmetric logistic model, the extremal dependence mixture structure means that collectively, $\lambda(\omega)$ and $\tau_i(\delta)$ $(i=1,2)$ do not provide sufficient information to deduce the shape of $G$. Hence, there is value in developing an approach to estimate $G$ directly, to enhance our understanding of the nature of the extremal dependence and to deduce self-consistent estimates of currently studied extremal dependence features. The remainder of the paper is dedicated to this goal.

\section{Estimation of the limit set $G$}\label{sec:method}
\subsection{Framework for modelling the boundary set $G$}\label{subsec:Gestimation}
For inference, we require the variables $(X_1,X_2)$ to have Exponential margins, which can be achieved in practice by applying rank and probability integral transforms to each variable. Suppose we have $n$ observations of the random variables $Y_i$, $i=1,2$, denoted by $y_{i,1},\dots,y_{i,n}$. The approximate transformation to observations of $X_i$ with Exponential margins is given by
\begin{equation}
x_{i,j} = -\log\left\{1-\frac{\text{rank}(y_{i,j})}{n+1}\right\},
\label{eqn:rankTransform}
\end{equation}
for $i=1,2$ and $j=1,\dots,n$. An alternative version of this transformation with a parametric form for the tail is detailed in \cite{ColesAndTawn1991}; this ultimately allows for marginal extrapolation on the original scale of the data, but where such extrapolation is not required (such as in our setting), the rank transform can lead to more robust estimates. Applying transformation~\eqref{eqn:rankTransform} as a preliminary step allows us to focus on estimating the extremal dependence features.

We consider $(X_1,X_2)$ in terms of the pseudo-polar coordinates $(R,W)$, with
\begin{equation}
    R = X_1+X_2 > 0, \qquad W=X_1/R\in[0,1],
\label{eqn:pseudo-polar}
\end{equation}
which we refer to as radial and angular components, respectively. Due to our choice of marginal distribution, these are not the same as the radial and angular components on Fr\'echet or Pareto scale used by \cite{Resnick1987} or \cite{ColesAndTawn1991}, for instance, but are akin to those in \cite{Wadsworth2013}. This polar coordinate formulation lends itself to estimating $G$ since the star-shaped nature of the set $G_S$ means that any ray emanating from the origin, i.e., corresponding to a fixed angle $w\in[0,1]$, must intersect the set $G$ exactly once. We are therefore interested in the largest possible radial value associated with each angle $w$, after the $\log n$ scaling of $(X_1,X_2)$, which also induces $\log n$ scaling of $R$. In practice, having a finite amount of data means that observations of $(X_1,X_2)/\log n$ are not restricted to the set $[0,1]^2$, unlike the set $G$ that we aim to estimate, and the distribution of the observed radial variable need not have a finite upper endpoint. We propose to estimate $G$ by estimating high quantiles of the distribution of $R\mid W=w$ for all $w\in[0,1]$, then transforming these estimates to the original coordinates $(X_1,X_2)$ via the relations $X_1=RW$, $X_2=R(1-W)$ and finally scaling them onto $[0,1]^2$. 

A natural candidate for modelling the tail of $R\mid W=w$ is the generalised Pareto distribution (GPD) \citep{Davison1990}. That is, for $w\in[0,1]$ and a suitable threshold $u_w$, we have
\begin{equation}
\Pr\left(R<r\mid R>u_w,W=w\right) = 1 -\left[1+\xi(w)\left\{\frac{r-u_w}{\sigma(w)}\right\}\right]^{-1/\xi(w)}_+,
\label{eqn:GPD}
\end{equation}
for $r>u_w$, $x_+=\max\{x,0\}$, scale parameter $\sigma(w)>0$ and shape parameter $\xi(w)\in\mathbbm{R}$. We consider $r_q(w)$, a high quantile of the distribution of $R\mid W=w$, for all $w\in[0,1]$, where 
\[
\Pr\{R<r_q(w)\mid W=w\}=q,
\]
with $r_q(w)>u_w$. For $\xi(w)\neq 0$ and $\zeta_u(w)=\Pr(R>u_w\mid W=w)$, this is 
\begin{equation}
r_q(w) = u_w + \frac{\sigma(w)}{\xi(w)}\left[\left\{\frac{\zeta_u(w)}{1-q}\right\}^{\xi(w)}-1\right].
\label{eqn:radialQuantile}
\end{equation} 
We aim to estimate the set $\{(r_q(w),w):w\in[0,1]\}$ for $q$ close to~1. 

\subsection{Overview of our inference procedure}\label{subsec:methodOverview}
In Sections~\ref{subsec:Gestimation2} and \ref{subsec:GestimationSmooth}, we propose a method to estimate the set $G$, by exploiting the framework detailed in Section~\ref{subsec:Gestimation}. In Section~\ref{subsec:Gestimation2}, a local estimation procedure is proposed, where estimates of the radial quantiles in~\eqref{eqn:radialQuantile} are obtained for a fixed set of angles by considering radial observations within some appropriate neighbourhood. We can induce an estimate of $G$ from these radial quantiles by applying an appropriate scaling onto the region $[0,1]^2$; this scaling procedure is discussed in Section~\ref{subsec:Gestimation2}. To achieve a smooth estimate of the set $G$, we then propose to exploit the generalised additive models (GAMs) framework for the GPD parameters in~\eqref{eqn:GPD}, with the degree of the corresponding splines informed by the results from the local quantile estimates. This is discussed in Section~\ref{subsec:GestimationSmooth} and provides our final estimate of $G$. An algorithmic representation of the full method is provided in the supplementary material \cite[Section L]{SupplementaryMaterial}.

\subsection{Estimation of $G$: a local approach}\label{subsec:Gestimation2}
Suppose we have $n$ observations of the radial-angular components $(R,W)$, denoted by $(r_1,w_1),\dots,$ $(r_n,w_n)$. If $(X_1,X_2)$ have a joint density, we will not observe repeated observations of the angular component in a finite sample, so for any given $w\in[0,1]$, we will not have sufficient observations\footnote{The rank transformation given in~\eqref{eqn:rankTransform} means that a small number of repetitions are possible.} to fit distribution~\eqref{eqn:GPD} directly. We propose a two-step approach to address this issue. First, we obtain local estimates of the model parameters $\sigma(w)$ and $\xi(w)$ and corresponding radial quantiles $r_q(w)$ for fixed values of $w$. From these, we construct a local estimate of the set $G$, which we denote $\hat G^L$. The theoretical examples in Figure~\ref{fig:gaugeEx} showed that the set $G$ may generally be smooth, but can have points of non-differentiability. Motivated by this observation, our second step, detailed in Section~\ref{subsec:GestimationSmooth}, exploits the GAMs framework to provide smooth estimates of the GPD threshold and scale parameter, and hence of $r_q(w)$, for $w\in[0,1]$. In Section~\ref{subsec:GestimationSmooth}, we explain how we use $\hat G^L$ to inform the choice of spline basis functions in the GAMs framework and yield an estimate $\hat G$ of $G$.

In the local estimation procedure, we use observed radial values corresponding to the angular components in some neighbourhood of $w$, defined as
\begin{equation}
\mathcal{R}_w = \left\{r_i:|w_i-w|\leq\epsilon_w, i\in\{1,\dots,n\}\right\},
\label{eqn:Rneighbours}
\end{equation}
for some $\epsilon_w>0$, to be equivalent to the event $W=w$, i.e., this requires the GPD parameters to be approximately constant over $\mathcal{R}_w$. Using standard maximum likelihood estimation techniques with observations from $\mathcal{R}_w$, it is straightforward to estimate the parameters $\sigma(w)$, $\xi(w)$ in~\eqref{eqn:GPD} and~\eqref{eqn:radialQuantile}, and we suggest setting $\zeta_u(w)=q_u\in[0,q]$ identically for all $w$ with the corresponding threshold $u_w$ estimated empirically. We propose to carry out this estimation at a range of angular values, denoted by $w^*_1,\dots,w^*_k$, and use these to estimate the radial quantiles in~\eqref{eqn:radialQuantile}, denoting these quantile estimates by $\hat{r}_q(w_j^*)$, $j\in J_k:=\{1,\dots,k\}$. The $w_j^*$ values are not required to belong to the set of observed $W$ values and may be selected so that the associated $\mathcal{R}_{w_j^*}$ sets overlap. As well as $k$ and the values $w^*_j$ $(j\in J_k)$, other tuning parameters to be selected here are the quantile levels $(q_u,q)$ and the neighbourhood sizes $\epsilon_{w^*_j}$. We discuss our approach to choosing these quantities in the supplementary material \cite[Section B]{SupplementaryMaterial}, where we propose choosing $\epsilon_{w^*_j}$ separately for each $j$ to ensure $|\mathcal{R}_{w^*_j}|=m$. Our default tuning parameter suggestions are $k=199$, $m=100$, $q_u=0.5$ and $q=0.999$.

To obtain an estimate of $G$, denoted $\hat G^L$ for the local approach, we first transform the local radial quantile estimates back to the original $(X_1,X_2)$ coordinates by setting
\begin{equation*}
\tilde{x}_{1,j} = \hat{r}_q(w_j^*)\cdot w^*_j, \qquad \tilde{x}_{2,j} = \hat{r}_q(w_j^*)\cdot (1-w_j^*),\qquad j\in J_k.
\end{equation*}
To ensure $G\subset[0,1]^2$, a naive approach would be to simply divide each component by its largest value, i.e., by updating to $\tilde{x}'_{i,j}=\tilde{x}_{i,j}/\max_{j\in J_k}\tilde{x}_{i,j}$, $i=1,2$. However, this is strongly influenced by the largest estimates, and does not work well in practice. Instead, we suggest a two-step scaling/truncation approach, initially informed by an existing estimator of $\eta$ and then corrected to take into account known features of $G$. We begin by scaling all $(\tilde{x}_{1,j},\tilde{x}_{2,j})$ values to obtain a preliminary local limit set estimate $\hat G_\eta^L=\left\{(\hat{x}^{G_\eta}_{1,j},\hat{x}^{G_\eta}_{2,j}):j\in J_k\right\}$, with the scaling factor chosen so that an estimate of $\eta$ from the resulting set would match the Hill estimator $\hat\eta_H$, see~\eqref{eqn:HillEta}, i.e., so that $\hat G^L_\eta$ and $[\hat\eta_H,\infty]^2$ just intersect. The scaling factor that achieves this is
\[
x^* = \hat\eta_H\left[\max_{j\in J_k}\left\{\min(\tilde{x}_{1,j},\tilde{x}_{2,j})\right\}\right]^{-1},
\]
i.e., we set $\hat x^{G_\eta}_{i,j}=x^*\tilde x_{i,j}$ for $i=1,2$, $j\in J_k$. The resulting estimate $\hat G^L_\eta$ is not guaranteed to lie within $[0,1]^2$, nor to intersect the lines $x_1=1$ and $x_2=1$; the second step corrects for these possibilities, taking each of the arguments separately. In step 2a, for $i=1,2$, if $\max_{j\in J_k}\hat x^{G_\eta}_{i,j}\geq 1$, we truncate by defining $\hat{x}^{G}_{i,j}=\min\left\{1,\hat{x}^{G_\eta}_{i,j}\right\}$, $j\in J_k$. Alternatively, in step 2b, if $\max_{j\in J_k}\hat x^{G_\eta}_{i,j}<1$, we ensure that our estimate of $G$ intersects the line $x_i=1$ at least once by dividing the component by its maximum value, setting $\hat{x}^{G}_{i,j}= \hat{x}^{G_\eta}_{i,j}/\max\limits_{\ell\in J_k}\hat{x}^{G_\eta}_{i,\ell}$. The scaling in step 2b means that an estimate of $\eta$ obtained from $\hat G^L$ (as will be discussed in Section~\ref{subsec:parameterEstimation}) may not be exactly equal to $\hat\eta_H$, but our approach actually improves the estimates of this coefficient when $\hat\eta_H$ has a tendency to underestimate, such as under asymptotic dependence in the logistic model; see Section~\ref{subsec:simulationStudy}. The full scaling/truncation approach is demonstrated pictorially in the supplementary material \cite[Section C]{SupplementaryMaterial}. 
This scaling procedure relies on us choosing an appropriate threshold $u_H$ for $\hat\eta_H$; we set this to the 0.95 quantile of the structure variable $M$ in~\eqref{eqn:HillEta}, which is also the level used throughout the simulation studies in Section~\ref{subsec:simulationStudy}, comparing different estimators of $\eta$.

\subsection{Estimation of $G$: a smoothed approach}\label{subsec:GestimationSmooth}
There is no guarantee of smoothness in the estimated set $\hat G^L$, whatever the choice of tuning parameters. As we have already identified, it may be desirable to impose that our estimator of $G$ is smooth, at least for some subsets of $w\in[0,1]$. To achieve this, we use $\hat G^L$ to inform a `smoothing' method for the GPD parameters in the second step of our procedure. In particular, we follow the approach of \cite{Youngman2019} by first carrying out quantile regression to estimate the thresholds $u_w$ at the same quantile level $q_u\in[0,q]$ as in Section~\ref{subsec:Gestimation2} by exploiting the asymmetric Laplace distribution \citep{Yu2001}. The parameters of the asymmetric Laplace distribution are modelled using GAMs. We find that applying this quantile regression approach to $R\mid (W=w)$ often gives impossible negative estimates near the largest and smallest observed angles. We overcome this issue by working with $\log R\mid (W=w)$ in this step, and then back-transform to obtain the required radial threshold estimates. For exceedances above $u_w$, we also assume that the parameter values in the GPD vary smoothly with the value of $w\in[0,1]$. Again, following the approach of \cite{Youngman2019}, we adopt a GAMs form for the log-scale parameter. Both parts of this approach are implemented in the \texttt{R} package \texttt{evgam} \citep{evgam}. The shape parameter could analogously be allowed to vary smoothly, but this is notoriously difficult to estimate and we found that in tests, anything other than assuming a constant value resulted in increased variance without reducing the bias of our estimates.

We allow the GAMs form for the model parameters to be linear, quadratic or cubic B-splines, with $\kappa$ knots; for simplicity, the same spline degree and knot positions are chosen for $u_w$ and $\log\sigma(w)$, and corresponding estimates of the quantiles in~\eqref{eqn:radialQuantile}. Once we have estimates of the high radial quantiles, the procedure to obtain an estimate of $G$ is via transformation and scaling/truncation, equivalent to that for $\hat G^L$. Although, theoretically, this approach allows for estimates of $G$ across the full range $w\in[0,1]$, outside the observed angles these can be unreliable, and we restrict our estimates to the observed range, $w\in(w^m,w^M)$. The resulting estimates of $G$ are denoted by $\hat G^{S_1}$, $\hat G^{S_2}$ and $\hat G^{S_3}$ for the linear, quadratic and cubic splines, respectively. 

\begin{figure}[t]
    \centering
    \includegraphics[width=11cm]{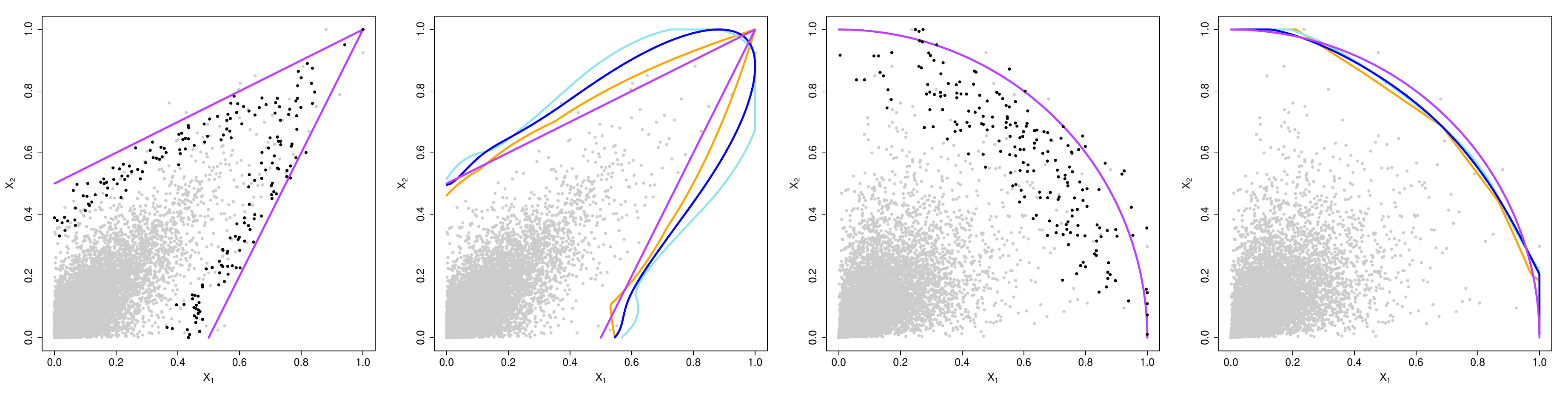} 
    \caption{Left: logistic data scaled by $\log n$, with the true set $G$ (purple) and $\hat G^L$ (black points). Left-centre: the estimated sets $\hat G^{S_1}$ (orange), $\hat G^{S_2}$ (light blue) and $\hat G^{S_3}$ (dark blue), taking $k=199$, $m=100$, $q_u=0.5$, $q=0.9$ and $\kappa=7$. Right-centre and right: equivalent plots for inverted logistic data.}
    \label{fig:Gexample2}
\end{figure}

Considering different spline degrees here is useful, as it enables us to capture different shapes of $G$. To choose between the estimators $\hat G^{S_1}$, $\hat G^{S_2}$ and $\hat G^{S_3}$, we select the spline degree where the estimated radial quantiles at the angles $w_j^*$ ($j\in J_k$) are closest to those from the local procedure, when compared using the mean absolute error. This selection criterion is a reasonable choice, since by forcing an equal number of points in each set $\mathcal{R}_{w_j^*}$, we have ensured there will be approximately the same uncertainty in each radial estimate. The selected estimator $\hat G$ (from $\hat G^{S_1}$, $\hat G^{S_2}$ or $\hat G^{S_3}$) becomes our preferred option for modelling $G$. That is, letting $\hat r^L_q(w^*_j)$, $j\in J_k$, denote the local radial quantile estimates and $\hat r^{S_d}_q(w^*_j)$, $d=1,2,3$, denote the corresponding estimates of $r_q(w^*_j)$ for $\hat G^{S_d}$, we have
\begin{equation}
\hat G = \left\{\hat G^{S_d}: d=\argmin_{d^*\in\{1,2,3\}}\sum_{j\in J_k}\left|\hat r^L_q(w^*_j)-\hat r^{S_{d^*}}_q(w^*_j)\right|\right\}.
\label{eqn:Gest}
\end{equation}

Figure~\ref{fig:Gexample2} gives a demonstration of the full approach for 10,000 simulated values from a bivariate extreme value model with a logistic model and parameter $\gamma=1/2$, and its inverted counterpart, with comparison to the true sets $G$. The tuning parameters here are chosen for illustrative purposes. It is clear that $\hat G^L$ is not smooth and tends to underestimate $G$, whereas $\hat G^{S_1}$, $\hat G^{S_2}$ and $\hat G^{S_3}$ offer varying levels of improvement. The linear spline performs the best in the logistic case where $G$ has a pointed shape, with the mean absolute errors evaluated in~\eqref{eqn:Gest} being 55.23, 67.37 and 57.64 for $\hat G^{S_1}$, $\hat G^{S_2}$ and $\hat G^{S_3}$, respectively, so $\hat G=\hat G^{S_1}$. The errors in the inverted logistic case are 61.95, 61.50 and 61.29, i.e., the cubic spline is slightly preferred as $G$ is smooth, so $\hat G=\hat G^{S_3}$. 

The smoothing method requires the introduction of further tuning parameters, in the choice of $\kappa$ and the position of the knots. Tuning parameter selection is discussed in detail in the supplementary material \cite[Section B]{SupplementaryMaterial}, where a default choice of $\kappa=7$ is proposed, in addition to the suggested tuning parameters for $\hat G^L$ in Section~\ref{subsec:Gestimation2}. Knot-placement; the positions of $w^*_j$ ($j\in J_k$); and sensitivity to the extrapolation quantile $q$ are also discussed in the supplementary material \cite[Sections B and E]{SupplementaryMaterial}, and we show limited sensitivity to $\kappa$ for an asymmetric logistic copula \cite[Section I]{SupplementaryMaterial}. 

\section{Estimation of extremal dependence properties}\label{sec:parameterEsts}
\subsection{Overview}
Further exploiting the new procedure to estimate $G$, in Section~\ref{subsec:parameterEstimation}, we obtain estimates of $\eta$, $\lambda(\omega)$, $\tau_1(\delta)$, $\tau_2(\delta)$, $\alpha_1$ and $\alpha_2$ in a unified way. Due to the subtlety of result~\eqref{eqn:NWbeta} for the conditional extremes parameters $\beta_1$ and $\beta_2$, we choose not to estimate this as part of our self-consistent approach. Instead, we fix the corresponding $\alpha_i$ parameter to its estimated value, and then carry out inference for $\beta_i$; this is discussed in Section~\ref{subsec:betaEstimation}. In Section~\ref{subsec:existingEstimators}, we detail some techniques that are currently available to separately estimate the extremal dependence features that we consider. In Section~\ref{subsec:properties}, we highlight some desirable properties of our estimators, including (but not limited to) the self-consistency that has been previously mentioned. We compare the performance of these methods to that of our estimation approach in Section~\ref{sec:results} and the supplementary material \cite[Sections F-H]{SupplementaryMaterial}. 

\subsection{Exploiting the estimate $\hat G$ for parameter estimation}\label{subsec:parameterEstimation}
For the estimation of the dependence properties, we consider $\hat G$ in~\eqref{eqn:Gest} at a finite number of angles, for ease of implementation. We choose these as identical to the $w_j^*$, $j\in J_k$, used in the local estimate $\hat G^L$, now denoting $\hat G=\left\{(\hat x_{1,j}^G,\hat x_{2,j}^G):j\in J_k\right\}$. To estimate $\eta$ using $\hat G$, we first note that result~\eqref{eqn:Neta} from \cite{Nolde2014} can be written as
\begin{equation}
\eta=\min\left\{s\in(0,1]: G\cap[s,\infty]^2=\emptyset\right\} = \max\left\{s\in(0,1]: G\cap[s,\infty]^2\not=\emptyset\right\}.
\label{eqn:Neta2}
\end{equation}
Considering each point in our set $\hat G$ to be a candidate for the intersection of $G$ and $[\eta,\infty]^2$ required in~\eqref{eqn:Neta2}, we can examine a corresponding set $[s^*_j,\infty]^2$ for each such points by setting $s^*_j=\min(\hat{x}^G_{1,j},\hat{x}^G_{2,j})$. Then the resulting estimate of $\eta$ is given by $\max(s^*_1,\dots,s^*_k)$, i.e.,
\begin{equation}
    \hat\eta_G = \max_{j\in J_k}\left\{\min \left(\hat{x}^G_{1,j},\hat{x}^G_{2,j}\right)\right\}.
\label{eqn:etahat}
\end{equation}
We estimate $\lambda(\omega)$, $\omega\in[0,1]$, via estimation of $s_\omega$ in~\eqref{eqn:NWlambda}. We have
\[
s_\omega = \min\left\{s\in[0,1]:sS_\omega\cap G =\emptyset\right\} = \max\left\{s\in[0,1]:sS_\omega\cap G \neq\emptyset\right\},
\]
with $S_\omega = \left\{(x_1,x_2): x_1\geq\omega/\max(\omega,1-\omega), x_2\geq(1-\omega)/\max(\omega,1-\omega)\right\}$.
Analogously to $\hat\eta_G$, we obtain candidate values \[
s^*_{\omega,j}=\max(\omega,1-\omega)\cdot\min\left\{\hat{x}^G_{1,j}/\omega,\hat{x}^G_{2,j}/(1-\omega)\right\}
\]
for $j\in J_k$, with the resulting estimator
\[
\hat\lambda_G(\omega)=\left[\max_{j\in J_k}\left\{\min\left(\frac{\hat{x}^G_{1,j}}{\omega},\frac{\hat{x}^G_{2,j}}{1-\omega}\right)\right\}\right]^{-1},\qquad w\in[0,1].
\]
For the dependence measure $\tau_1(\delta)$, result~\eqref{eqn:NWtau} gives that for $\delta\in[0,1]$,
\begin{align*}
\tau_1(\delta) &= \min\left\{s\in[0,1]:[s,\infty]\times[0,\delta s]\cap G =\emptyset\right\}\\
&= \max\left\{s\in[0,1]:[s,\infty]\times[0,\delta s]\cap G \not=\emptyset\right\}.
\end{align*}
Our candidate points for the intersection of the sets $G$ and $(\tau_1(\delta),\infty]\times[0,\delta\tau_1(\delta)]$ are all our estimated points on $G$ where $\hat x_2^G\leq\delta \hat x_1^G$. Our estimate of $\tau_1(\delta)$ corresponds to the largest such value of $\hat x_1^G$. That is,
\[
\hat\tau_{G,1}(\delta) = \max\left(\hat{x}^G_{1,j} : \hat{x}^G_{2,j} \leq \delta\hat{x}^G_{1,j}, j\in\{1,\dots,k\} \right).
\]
Finally, for the conditional extremes parameter $\alpha_1$, 
\begin{equation}
    \hat\alpha_{G,1} = \max\left(\hat x_{2,j}^G:\hat x_{1,j}^G=1, j\in\{1,\dots,k\}\right).
    \label{eqn:alphaEst}
\end{equation}
Due to our method of scaling onto $[0,1]^2$, in practice there may only be exactly one value of $\hat x_{1,j}^G$ that is equal to one. The estimators of $\hat\tau_{G,2}(\delta)$ and $\hat\alpha_{G,2}$ are defined analogously. Our approach ensures self-consistency across the conditioning variables in the conditional extremes approach, since $\hat\alpha_{G,1}=1$ if and only if $\hat\alpha_{G,2}=1$; this issue is also discussed in \cite{Liu2014}. Moreover, estimating the $\alpha_i~(i=1,2)$ parameters in this way avoids the need to place a distributional assumption on the random variable $Z$ in~\eqref{eqn:HandT}; this benefit over other approaches is discussed further in Section~\ref{subsec:properties}. A pictorial illustration of the estimation of $\eta$, $\lambda(\omega)$, $\tau_1(\delta)$ and $\alpha_1$ is given in the supplementary material \cite[Section C]{SupplementaryMaterial}.

\subsection{Estimation of the conditional scaling parameter $\beta_i$}\label{subsec:betaEstimation}
As discussed in Section~\ref{sec:NWtheory}, the parameter $\beta_i$, $i=1,2$, in the conditional extremes framework is also linked to the asymptotic shape of a scaled sample cloud. However, the required feature of the set $G$ is very subtle, and reliably estimating $\beta_i$ using our estimate $\hat G$ is not really feasible. Instead, we estimate $\beta_i$ with $\alpha_i$ fixed to its estimated value, $\hat\alpha_{G,i}$ of expression \eqref{eqn:alphaEst}. Illustrating this for $\beta_1$, limit~\eqref{eqn:HandT} is taken as an equality for some finite, but large, threshold $u$. Then, for $X_1=x>u$ and with $a_1(x)=\hat\alpha_{G,1} x$ and $b_1(x)=x^{\beta_1}$, this implies
\begin{equation*}
    X_2 \vert (X_1=x) = \hat\alpha_{G,1} x + x^{\beta_1} Z, \qquad x>u,
\end{equation*}
where $Z$ is a non-degenerate random variable that is independent of $X_1$. It is common to take $Z\sim\mathcal{N}(\mu,\sigma^2)$ as a working assumption, for $\mu\in\mathbbm{R}$, $\sigma>0$ (e.g., \cite{Keef.al.2013}), although other approaches have been considered (e.g., \cite{Lugrin2016}), which implies 
\begin{equation}
    X_2 \vert (X_1=x) \sim \mathcal{N}\left\{(\hat\alpha_{G,1} x + x^{\beta_1} \mu),(x^{\beta_1}\sigma)^2\right\}, \qquad x>u.
    \label{eqn:HTestimate}
\end{equation}
From this, it is straightforward to obtain maximum likelihood estimates for the parameters, with $\hat\beta_{G,1}$ denoting the corresponding estimate for $\beta_1$. 

\subsection{Existing estimation techniques}\label{subsec:existingEstimators}
In this section, we briefly present some existing estimators for the extremal dependence properties discussed in Section~\ref{subsec:BEVfeatures}. We use these for comparison to our proposed estimators in Section~\ref{sec:results}. More detail on these estimators is provided in the supplementary material \cite[Section D]{SupplementaryMaterial}.

For $\eta$, we consider three options. The first is a Hill-type estimator \citep{Hill1975} on the structure variable $M=\min(X_1,X_2)$. Assuming we have $n_{u,H}$ observations of $M$ above some high threshold $u_H$, denoted $m^*_1,\dots,m^*_{n_{u,H}}$, the estimator is 
\begin{equation}
\hat\eta_{H} = \min\left\{\frac{1}{n_{u,H}}\sum\limits_{i=1}^{n_{u,H}} \left(m^*_i-u_H\right),1\right\}.
\label{eqn:HillEta}    
\end{equation}
Alternatively, suppose we have $n$ pairs of observations $(x_{1,1},x_{1,2}),\dots,(x_{n,1},x_{n,2})$, and denote by $x^{(j)}_i$ the $j$th largest value of component $i\in\{1,2\}$, i.e., $x_i^{(n)}\leq x_i^{(n-1)}\leq\dots\leq x_i^{(1)}$. Then for each $j\in\{1,\dots,n\}$, define the quantity $s_n(j) = \sum_{\ell=1}^n \bm{1}\left\{x_{\ell,1} \geq x^{(j)}_1, x_{\ell,2} \geq x^{(j)}_2\right\}$. Two non-parametric estimators, $\hat\eta_P$ of \cite{Peng1999} and $\hat\eta_D$ of \cite{Draisma2004}, are
\begin{equation*}
    \hat\eta_{P} = \min\left[\frac{\log 2}{\log\left\{s_n(2c)\right\}-\log\left\{s_n(c)\right\}},1\right], ~~
    \hat\eta_D = \min\left\{\frac{\sum_{j=1}^cs_n(j)}{cs_n(c)- \sum_{j=1}^cs_n(j)},1\right\},
\end{equation*}
where the forms of these estimators are as presented in \cite[p.352]{Beirlant2004}, but we have added the truncation at one. The tuning parameter $c$ relates to the number of exceedances above a high threshold.

Hill-type estimators are also available for $\lambda(\omega)$ and $\tau_1(\delta)$. For the former, set $M_\omega=\min\left\{X_1/\omega,X_2/(1-\omega)\right\}$, for $\omega\in[0,1]$, and assume there are $n_{u,\omega}$ observations of $M_\omega$ above a high threshold $u_\omega$, denoted $m^*_{\omega,1},\dots,m^*_{\omega,n_{u,\omega}}$. The corresponding estimator is
\begin{equation*}
\hat\lambda_{H}(\omega) = \min\left[\left\{\frac{1}{n_{u,\omega}}\sum\limits_{i=1}^{n_{u,\omega}} \left(m^*_{\omega,i}-u_\omega\right)\right\}^{-1}, 1\right].
\end{equation*}
For $\tau_1(\delta)$, $\delta\in(0,1]$, let $M_\delta = \{X_1>0 : X_2 \leq \delta X_1\}$ and suppose we have $n_\delta\leq n$ observations of $X_1$ where $X_2 \leq \delta X_1$, and that $n_{u,\delta}$ of these, denoted $m^*_{\delta,1},\dots, m^*_{\delta,n_{u,\delta}}$, are above some high threshold $u_\delta$. Then the estimator is 
\begin{equation*}
\hat\tau_{H,1}(\delta) = \min\left\{\frac{1}{n_{u,\delta}}\sum\limits_{i=1}^{n_{u,\delta}} \left(m^*_{\delta,i}-u_\delta\right), 1\right\}.
\end{equation*}

\subsection{Properties of our estimators}\label{subsec:properties}
The main advantage of our approach over estimating each of the extremal dependence features separately, is that by linking each of the estimators to a single estimate of $G$, we automatically achieve self-consistency across the different dependence features. One way to think about this is in terms of whether the parameter estimates indicate asymptotic dependence or asymptotic independence, since separate estimation could lead to contradictions. Returning to our example from Section~\ref{sec:intro}, we consider the case of asymptotic dependence, where we know that $\eta=1$ and $\alpha_1=\alpha_2=1$. In our approach, we obtain $\hat\eta_G=1$ and $\hat\alpha_{G,1}=\hat\alpha_{G,2}=1$ exactly when our estimate of the set $G$ contains $(1,1)$. This means that $\hat\eta_G=1$ if and only if $\hat\alpha_{G,1}=\hat\alpha_{G,2}=1$, which is not guaranteed when carrying out inference separately for these dependence features.

We mentioned in Section~\ref{subsec:BEVfeatures} that the set of measures $\{\eta,\tau_1(\delta),\tau_2(\delta)\}$ can be used to describe the extremal dependence structure of the variables. In particular, if $X_1$ can be large while $X_2$ is small, there must be some value of $\delta^*<1$ such that $\tau_1(\delta^*)=1$, and a similar result holds for $\tau_2(\delta)$. Moreover, if both variables can take their largest values simultaneously, then $\eta=1$. Simpson et al.\ \cite{Simpson2020} point out that
\begin{equation}
    \max\left\{\eta, \tau_1(\delta): \delta\in[0,1)\right\} = \max\left\{\eta, \tau_2(\delta): \delta\in[0,1)\right\} = 1.
\label{eqn:tauEtaConstraint}
\end{equation}
This is not guaranteed by $\hat\eta_H$, $\hat\tau_{H,1}(\delta)$ and $\hat\tau_{H,2}(\delta)$ (although a possible way to enforce this is discussed in \cite{Simpson2019}), but is naturally satisfied by our new estimation procedure, since the scaling/truncation steps ensure $\hat G$ intersects both the lines $x_1=1$ and $x_2=1$ at least once. Further, we are guaranteed to have monotonicity in our estimates of $\tau_i(\delta)~(i=1,2)$ as $\delta$ increases; this theoretical property is not achieved through application of the Hill estimator at separate values of $\delta$.

Finally, maximum likelihood estimation for conditional extremes models involves placing a distributional assumption on $Z$. This often gives reasonable performance, but mis-specifying the distribution has the potential to affect the resulting estimates of $\alpha_i$ and $\beta_i$. Our new approach has the benefit that we do not need to place a distributional assumption on $Z$ for estimation of $\alpha_i$, although we do for subsequent estimation of $\beta_i$.

\section{Results - simulation study and data example}\label{sec:results}
\subsection{Simulation study}\label{subsec:simulationStudy}
We now demonstrate the performance of our proposed estimation procedure via a simulation study. Our primary interest lies in estimating the whole of $G$, which is the focus of Section~\ref{subsec:simulationStudyG}. However, no other estimators of $G$ exist, but there are estimators available for certain parts of $G$ if we exploit the links to the extremal dependence measures discussed in Sections~\ref{sec:theory}~and~\ref{sec:parameterEsts}. In order to judge our performance against these competitors, we provide additional simulation results for individual dependence measures in Section~\ref{subsec:simulationStudyParams}. Our estimation procedure provides point estimates for $G$ and each of the individual dependence measures; to assess uncertainty, we propose that a non-parametric bootstrapping procedure could be used, and show in the supplementary material \cite[Section K]{SupplementaryMaterial} that this approach can provide reasonable coverage.

\subsubsection{Estimation of $G$}\label{subsec:simulationStudyG}
\begin{figure}[t]
    \centering
    \includegraphics[width=11cm]{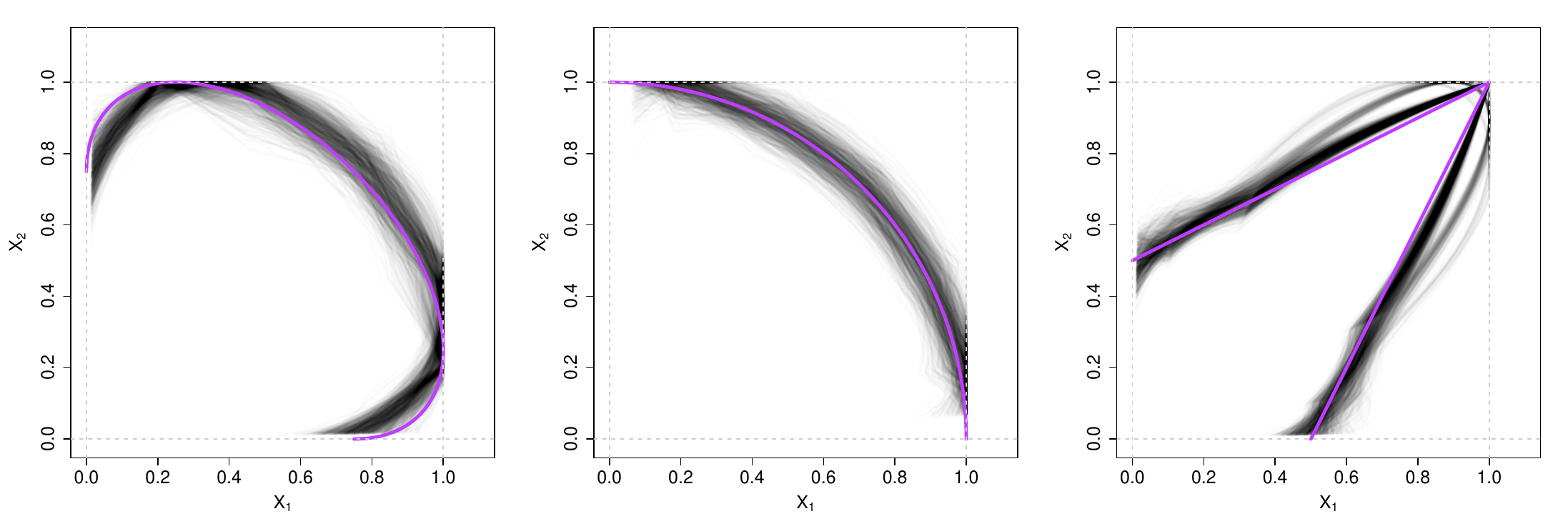}  
    \caption{Estimates $\hat G$ for the Gaussian (left), inverted logistic (centre) and logistic (right) models, with the $\rho$ or $\gamma$ parameters set to 0.5. The sample size in each case is $n=10,000$ and 1000 estimates are shown for each model (grey). The true sets $G$ are shown in purple.}
    \label{fig:Gexample3}
\end{figure}

We begin by considering estimation of the complete set $G$ for the first three copulas from Figure~\ref{fig:gaugeEx}, with $\rho,\gamma=0.5$. For each copula, with Exponential margins, we take a sample of size $10,000$ and use the estimator $\hat G$ of~\eqref{eqn:Gest}. This is repeated 1000 times for each model, with the resulting estimates $\hat G$ shown in Figure~\ref{fig:Gexample3}, alongside the true $G$. To obtain these results, we select the tuning parameters as stated in Section~\ref{subsec:GestimationSmooth}. In the supplementary material \cite[Section J]{SupplementaryMaterial}, we demonstrate the performance of our $\hat G$ for samples of size $n=100,000$ taken from the same three copula models; these results demonstrate that the tuning parameter choices are also reasonable for larger samples.

\begin{table}[t]
    \centering
    \begin{tabular}{c|ccc}
        Spline degree &  Gaussian  &  Inverted logistic  &  Logistic \\
        \hline 
        1 &  392  &  355  &  798  \\
        2 &  482  &  250  &   32  \\
        3 &  126  &  395  &  170  \\
    \end{tabular}
    \caption{The number of times each spline degree is chosen for the sets $\hat G$ shown in Figure~\ref{fig:Gexample3}.}
    \label{tab:splineResults}
\end{table}

There is reasonable agreement between $\hat G$ and $G$ in Figure~\ref{fig:Gexample3} for all three copulas. In the inverted logistic simulations, the shape of $G$ is generally well represented by $\hat G$. Some small bias occurs in the Gaussian case; this is investigated further in the supplementary material \cite[Section J]{SupplementaryMaterial}, where we discuss known issues with slow convergence for this copula. For the logistic case, there are some estimates where the pointed shape of $G$ has not been captured, but the estimates are reasonably successful overall. 

The degree of the spline chosen to model the log-scale parameter and threshold in the GPD plays an important role in determining the shape of $\hat G$; in Table~\ref{tab:splineResults}, we summarise the selected spline degrees over the 1000 replicated datasets for each copula. Since $G$ is smooth for both the Gaussian and inverted logistic copulas, we would expect to see the procedure favouring quadratic or cubic splines. For both these copulas, there is no overwhelming favourite in terms of the spline degrees chosen, but the non-linear options are selected around 60-65\% of the time. On the other hand, the pointed shape of $G$ for the logistic copula suggests that linear splines should be preferred here, with this being achieved in almost 80\% of cases, and $(1,1)\not\in \hat G$ corresponding to the selection of quadratic or cubic splines.

\subsubsection{Estimation of the dependence measures}\label{subsec:simulationStudyParams}
\begin{figure}[t]
    \centering
    \includegraphics[width=11cm]{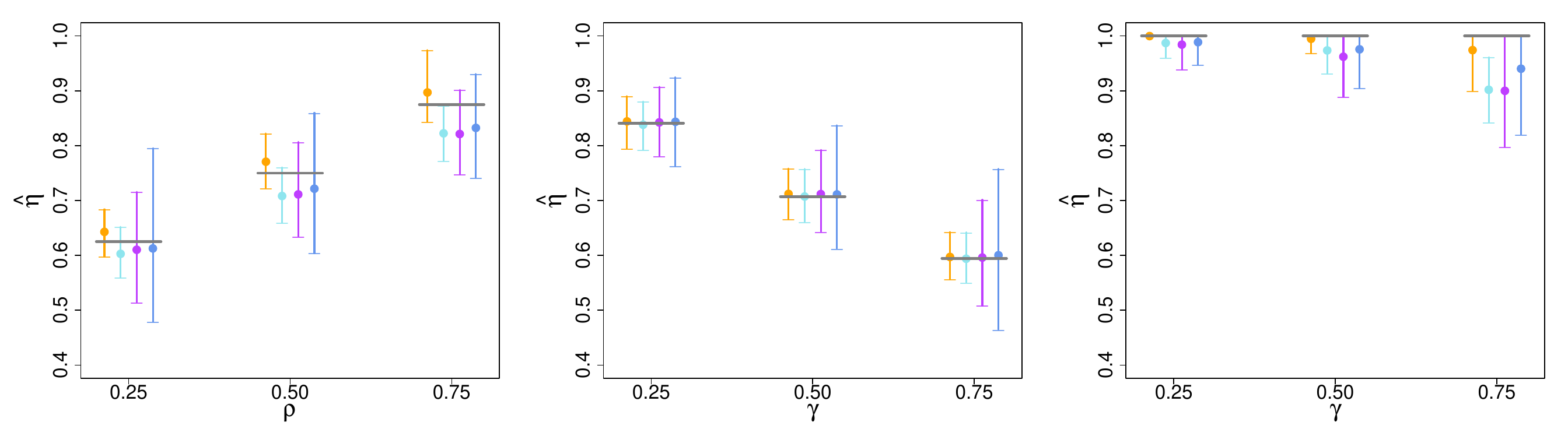} 
    \caption{Estimates of $\eta$ for data simulated from Gaussian (left), inverted logistic (centre) and logistic (right) copulas with their corresponding parameters taken as 0.25, 0.5 and 0.75. We show the mean estimates of $\eta$ (circles) and 0.025 to 0.975 quantile range from 1000 replicates for $\hat\eta_G$ (orange), $\hat\eta_H$ (light blue), $\hat\eta_P$ (purple), $\hat\eta_D$ (dark blue), each based on a data sample of 10,000. The true $\eta$ values, summarised in Table~\ref{tab:parameterValues}, are shown in grey. }
    \label{fig:etaEstimates}
\end{figure}

As discussed in Section~\ref{subsec:parameterEstimation}, once we have $\hat G$, we can use this to induce estimates of various extremal dependence properties. For estimating $\eta$, we compare our $\hat\eta_G$ given in~\eqref{eqn:etahat}, with the existing estimators $\hat\eta_H$, $\hat\eta_P$ and $\hat\eta_D$ described in Section~\ref{subsec:existingEstimators}. We fix the threshold $u_H$ used in the estimator $\hat\eta_H$ to coincide with $c$ for the estimators $\hat\eta_P$ and $\hat\eta_D$, such that all estimators use 500 data points. As with the tuning parameters in our approach, we have not optimised these choices, but have selected values that we found to work reasonably well across a range of examples. We take the set-up of our study as the same as for the results in Figure~\ref{fig:Gexample3}, but now take $\rho,\gamma\in\{0.25, 0.5, 0.75\}$; summaries of the $\eta$ estimates are shown in Figure~\ref{fig:etaEstimates}.

For the Gaussian copula, the existing approaches underestimate the true value of $\eta$, while our approach tends to overestimate by a similar magnitude, although our 95\%  sampling distribution interval does contain the truth in each case and is generally the narrowest. For the inverted logistic copula, both $\hat\eta_G$ and $\hat\eta_H$ perform consistently well across the different values of $\gamma$; all estimators appear to be unbiased but $\hat\eta_P$ and $\hat\eta_D$ suffer from increasing variability. The similarity of $\hat\eta_G$ and $\hat\eta_H$ is due to the fact that for this copula, step 2b of our scaling/truncation procedure is generally not required. Where $\hat\eta_G$ is particularly successful is in the logistic case, with asymptotic dependence and $\eta=1$. Across all values of $\gamma$, we outperform the existing estimators, with clear success in Figure~\ref{fig:etaEstimates} for $\gamma\in\{0.25,0.5\}$, where the dependence is stronger. For $\gamma=0.75$, the picture is less clear, so we find the root mean square error for each of the estimators in this case; these are 0.044, 0.103, 0.113 and 0.080 for $\hat\eta_G$, $\hat\eta_H$, $\hat\eta_P$ and $\hat\eta_D$, respectively, confirming that $\hat\eta_G$ provides improvement over previous approaches. In terms of self-consistency for the logistic copula, while our approach satisfies the required property that $\eta\geq\max(\alpha_1,\alpha_2)$ in all cases, this is only true for 45\%, 46\% and 66\% of the $\hat\eta_H$, $\hat\eta_P$ and $\hat\eta_D$ estimates in Figure~\ref{fig:etaEstimates}, respectively, when compared to the maximum likelihood estimator for $\alpha_i$, $i=1,2$ (results for which are given in the supplementary material \cite[Section H]{SupplementaryMaterial}). The existing estimators are almost always self-consistent in our study for the Gaussian and inverted logistic copulas, where $\eta$ is much larger than $\alpha_1=\alpha_2$, but our approach still has the benefit that this coherence is guaranteed.

\begin{figure}[t]
    \centering
    \includegraphics[width=11cm]{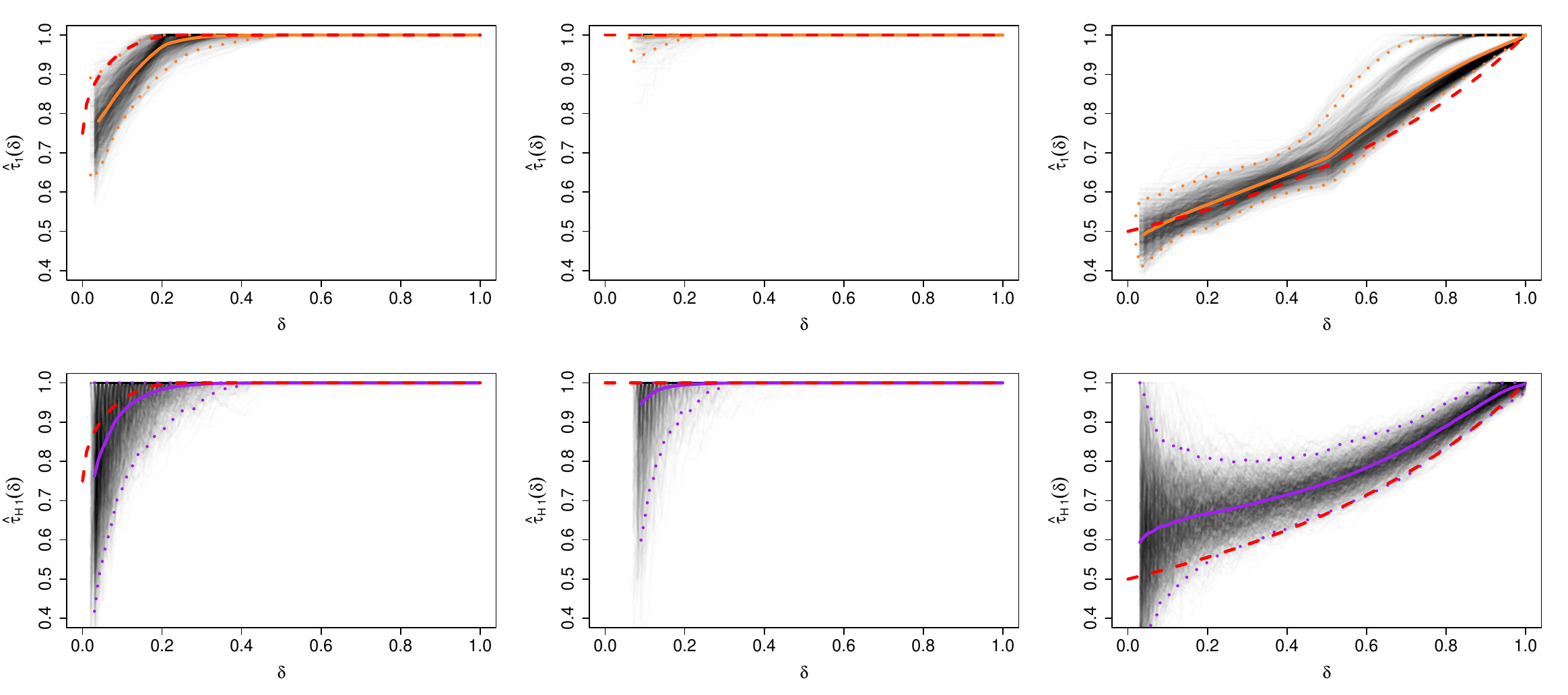}
    \caption{Estimates of $\tau_1(\delta)$, $\delta\in[0,1]$, for data simulated from Gaussian (left), inverted logistic (centre) and logistic (right) models with their corresponding parameters set as 0.5. Each grey line represents an estimate obtained using $\hat\tau_{G,1}(\delta)$ (top) or $\hat\tau_{H,1}(\delta)$ (bottom) over 1000 simulations. The solid and lower/upper dotted lines show the pointwise bootstrapped means and 0.025 and 0.975 quantiles, respectively, for $\hat\tau_{G,1}(\delta)$ (orange) and $\hat\tau_{H,1}(\delta)$ (purple). The true $\tau_1(\delta)$ values, summarised in Table~\ref{tab:parameterValues}, are shown in red.}
    \label{fig:tauEstimates0.5}
\end{figure}

In Figure~\ref{fig:tauEstimates0.5}, we present estimates of $\tau_1(\delta)$, comparing our procedure to $\hat\tau_{H,1}(\delta)$, described in Section~\ref{subsec:existingEstimators}. Following \cite{Simpson2020}, we fix the threshold $u_{\delta}$ 
for $\hat\tau_{H,1}(\delta)$ to the observed 0.85 quantile of the variable $M_\delta$. Since we restrict $\hat G$ to values within the range of observed angular values, there are some values of $\delta$ for which we cannot compute $\hat\tau_{G,1}(\delta)$. To ensure a fair comparison, we present the estimates $\hat\tau_{H,1}(\delta)$ over the same range of $\delta$. In terms of bias, the two procedures are relatively similar, but there are two features to highlight in our estimates. First, only 18.8\% of the $\hat\tau_{H,1}(\delta)$ estimates in Figure~\ref{fig:tauEstimates0.5} (measured over 0.01 increments of $\delta$) satisfy the required monotonicity property of $\tau_1(\delta)$, while our approach guarantees this aspect of self-consistency across different $\delta$ values in 100\% of cases. Second, for small values of $\delta$, $\hat\tau_{H,1}(\delta)$ has much greater variability than  $\hat\tau_{G,1}(\delta)$ for all three copulas with $\rho,\gamma=0.5$. This arises as $\hat\tau_{G,1}(\delta)$ `borrows' more information from nearby values of $\delta$ than  $\hat\tau_{H,1}(\delta)$, and therefore suffers less from the lack of data associated with estimates for small $\delta$ values.

In the supplementary material \cite[Section F]{SupplementaryMaterial}, we give further results for $\tau_1(\delta)$ for the copulas used in Figure~\ref{fig:tauEstimates0.5} with $\rho, \gamma\in\{0.25,0.75\}$; we see similar improvements in the logistic case with $\gamma=0.25$. We also provide results for estimation of the features $\lambda(\omega)$, $\alpha_1$ and $\beta_1$. Both $\hat\lambda_G(\omega)$ and $\hat\lambda_H(\omega)$ provide successful estimates for the inverted logistic model across a range of dependence parameters $\gamma$, with $\hat\lambda_G(\omega)$ providing smoother estimates; our approach exhibits less bias than $\hat\lambda_H(\omega)$ in Gaussian cases; and we see less variability in $\hat\lambda_G(\omega)$ than $\hat\lambda_H(\omega)$ in results for the logistic model. Our estimates of $(\alpha_1,\beta_1)$ are reasonably similar to those obtained by using maximum likelihood estimation for both parameters $\alpha_1$ and $\beta_1$ simultaneously, i.e., replacing $\hat\alpha_{G,1}$ in~\eqref{eqn:HTestimate} with $\alpha_1$ to be estimated. We also provide additional simulation results for the asymmetric logistic model.

We acknowledge that, considering estimates of each of the dependence measures separately, our new approach has some relatively small improvements over existing methods. However, we emphasise again the important point that our estimates are jointly self-consistent, which is not generally the case with the other approaches.

\subsection{Application to sea wave height data}\label{subsec:waves}
We now apply our approach to hindcast values of significant wave heights for locations in the North Sea. These are a subset of the data studied by \cite{Wadsworth2012} and, as in their paper, we consider only data from the winter months of December to February to limit issues with non-stationarity. We take data from four locations lying on a transect from the north-west to the south-east. The data cover 31 years, with eight observations for each day in our months of interest, resulting in a total of 22,376 observations per location. The transformation~\eqref{eqn:rankTransform} is applied to each variable to ensure standard Exponential marginal distributions.

\begin{figure}[t]
    \centering
    \includegraphics[width=5cm]{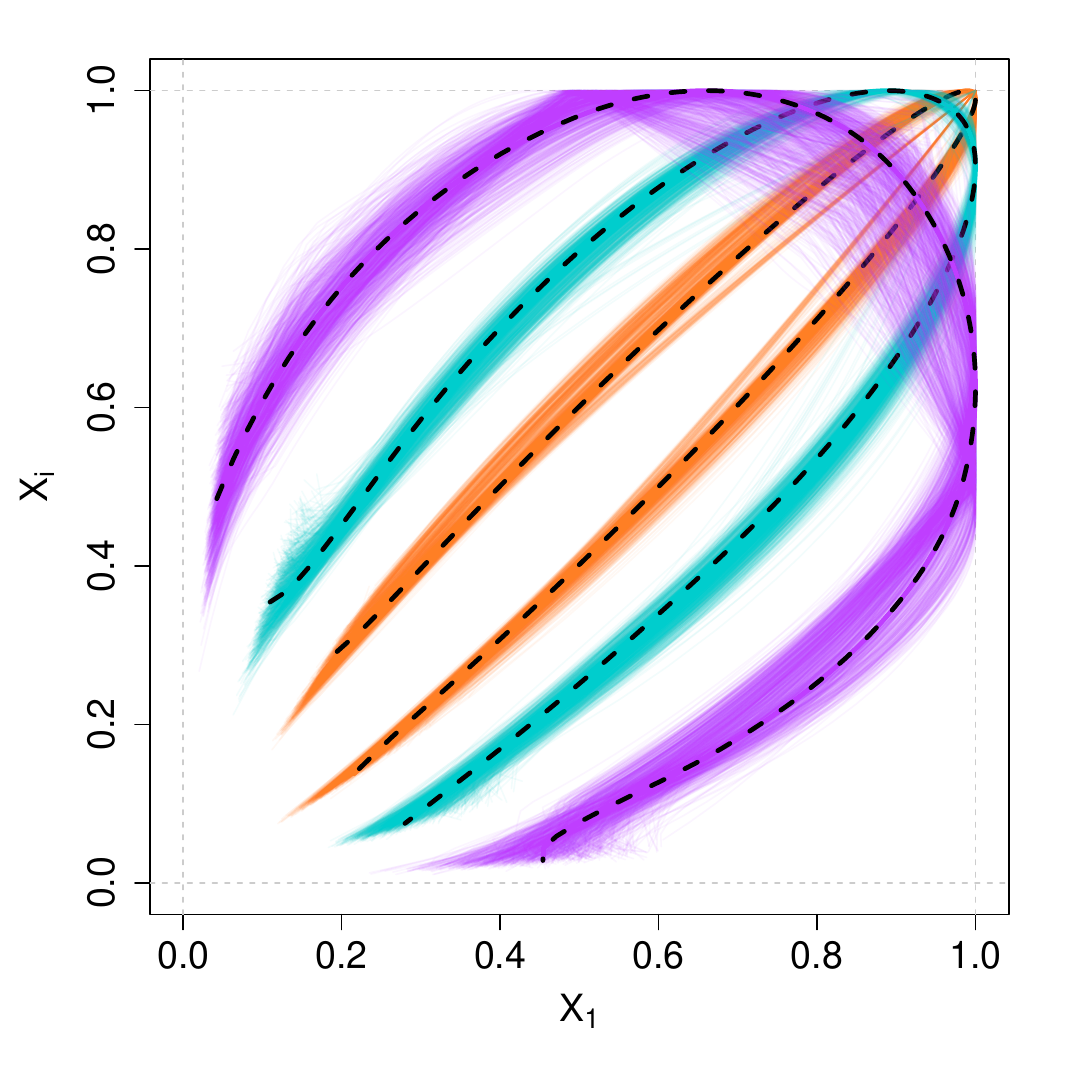}
    \caption{Estimates of $G$ for significant wave heights at three pairs of locations in the North Sea (black dashed lines) with 1000 bootstrapped estimates shown by the orange, turquoise and purple lines for locations at distances of approximately 40km, 130km, 300km, respectively. The variable $X_1$ represents the transformed and scaled significant wave heights at a location in the north-west of our study region; $X_i$ ($i=2,3,4$) corresponds to significant wave heights at the other locations.}
    \label{fig:waveResults}
\end{figure}

Figure~\ref{fig:waveResults} shows $\hat G$ at three pairs of locations with varying distances between them, with the tuning parameters as in Sections~\ref{subsec:GestimationSmooth} and~\ref{subsec:simulationStudy}. Uncertainty is presented by implementing the stationary bootstrap scheme of \cite{PolitisRomano1994} over 1000 iterations with the mean block-length set to 16 observations, i.e., two days. The most noticeable feature in these results is the weakening dependence in the significant wave height data as the distance between the locations increases, as is expected in environmental settings. The uncertainty in our estimates increases with the distance between the locations. While previous methods could have captured the weakening dependence for single parameter values, we can now see this change over the whole shape of $\hat G$, highlighting the value added by our approach. In terms of risk, these results can be interpreted based on whether pairs of locations may see extreme events simultaneously, compounding the potential impact. For the closest pair of locations, there is a clear risk that the largest wave heights can occur together. However, as the distance between locations increases, the largest wave heights at one location are associated with increasingly smaller events at the other location, and there becomes less risk of a severe impact at these two sites simultaneously.

For the closest pair of locations, we estimate the coefficient of tail dependence as $\hat\eta_G=0.997$ with a 95\% confidence interval of $[0.994, 1.000]$; for comparison, we estimate $\hat\eta_H=0.979$ $[0.962, 0.997]$. Our $\hat\eta_G$ suggests selecting a model that can capture asymptotic dependence would be most appropriate here, whereas using $\hat\eta_H$ the choice is more ambiguous due to a wider confidence interval that does not contain one. For the remaining two pairs of locations, the estimates are $\hat\eta_G=0.974$ $[0.964, 0.983]$ and $\hat\eta_G=0.897$ $[0.811, 0.932]$, respectively, with corresponding estimates $\hat\eta_H=0.918$ $[0.879, 0.954]$ and $\hat\eta_H=0.801$ $[0.748, 0.855]$. The $\hat\eta_G$ results both indicate that a model for asymptotic independence would be reasonable, although one may prefer the conservative approach of choosing an asymptotically dependent model for the former since the confidence interval is quite close to one. For all three pairs of locations, $\hat\eta_G>\hat\eta_H$, with narrower confidence intervals obtained for $\hat\eta_G$.

\section{Discussion}\label{sec:discussion}

\subsection{Summary of our contributions}
The aim of this paper was to develop a new inferential technique for bivariate extremes, providing estimation for a limit set and, as a by-product, yielding self-consistent estimation across a range of extremal dependence features. This was motivated by theoretical developments from \cite{Nolde2014} and \cite{Nolde2021}, who showed that these features can all be linked to the asymptotic shape of a suitably-scaled sample cloud. We have demonstrated that our approach provides estimators which, when considered marginally, are competitive compared to currently available techniques,
and provide particularly good improvement for the important case of asymptotic dependence. But critically, in terms of self-consistency, our results are highly superior, as highlighted in Section~\ref{subsec:properties}, and our method provides additional information about joint extreme events which is not captured by any existing extremal dependence measure, as most clearly illustrated in the discussion of Figure~\ref{fig:mixtureExamples} in Section~\ref{subsec:issues}.

\subsection{Links to other approaches}
The boundary function may appear at first sight to be essentially like a contour of equal joint density,  however, they are different quantities, which is highlighted by the examples in Section~\ref{subsec:Examples}. Also, the existing estimators of density contours in the tail of the joint distribution have major restrictions for practical use, which are not an issue in our consideration of $G$. Specifically, \cite{Cai2011, Einmahl2013} assume that a multivariate regular variation limit forms hold, which imposes that the marginal distributions must be heavy tailed, and that the variables are either asymptotically dependent or completely independent, meaning that more general forms of asymptotic independence are excluded. Furthermore, they impose conditions such that the spectral measure of the regular variation is non-zero across all interior directional rays.

While our paper has been in the review process, the draft version with working title \emph{``Estimating the limiting shape of bivariate scaled sample clouds for self-consistent inference of extremal dependence properties''} that was uploaded to \emph{arXiv} has been referenced in several subsequent papers also aiming to estimate the limit set, highlighting the value in developing such estimators; see for example \cite{Majumder2023, MurphyBarltrop2023, Papastathopoulos2023}. As part of their paper, Murphy-Barltrop et al.\ \cite{MurphyBarltrop2023} carry out an in-depth simulation study comparing estimators of $\lambda(\omega)$. One of their main messages is that ``global estimators'' such as ours, where simultaneous estimation of $\lambda(\omega)$ is carried out across all values of $\omega\in[0,1]$, far out-perform local estimates, and the approach we have proposed in particular is shown to perform very well. Moreover, Wadsworth and Campbell \cite{WadsworthCampbell2022} have independently developed what they term a ``geometric approach'' to estimating the probability of multivariate extreme events, with links to the framework we have presented here, and showing benefits over existing methods for multivariate extremes. Part of their inference involves estimation of the gauge function, which they restrict to a class of parametric forms; there is the potential for our proposed method for estimating $G$ to be combined with their approach for enhanced flexibility in this aspect. Overall, statistical inference with the limit set and gauge functions appears to be a promising avenue for future research.

An intermediate step of our estimation procedure considers radial quantiles at high (but non-limiting) levels. Considering these radial quantiles across the range of angles gives a contour that no longer represents a limit set, but which is still a potentially interesting extreme set in its own right. In particular, this can be linked to the idea of environmental contours, often used in engineering applications, and the methodology presented in this paper has now been extended to such a setting \cite{Simpson2024}.

\subsection{Potential issues and extensions}\label{subsec:issues}
We have assumed that our variables have standard Exponential margins. This restricts us to cases where there is independence or positive dependence between the variables, i.e., we cannot handle negatively dependent variables. For conditional extremes modelling, Keef et al.\ \cite{Keef.al.2013} capture this feature by instead using Laplace margins; this possibility is also briefly discussed by \cite{Nolde2021}. In practice, one could first test for negative dependence between the variables, and proceed with our approach only if it is not present in the data. It would also be possible to extend the methodology presented here for data initially transformed to Laplace margins, with different forms for the (pseudo-)polar coordinates chosen to allow the angular variable to cover all four quadrants, i.e., $W\in[-\pi,\pi)$; this approach is taken for the environmental contours in \cite{Simpson2024}.

As discussed in Section~\ref{sec:intro}, the limit set $G$ could be linked to likely boundary sets for large finite samples, which would take the form $G_m$ in \eqref{eqn:finiteSampleG} on Exponential margins. However, there may also be cases where a boundary set is required for finite samples on the original scale of the data. Taking the original variables as $(Y_1,Y_2)$ with marginal distributions $F_{Y_{i}}(y)$ $(i=1,2)$, one could achieve this by considering 
\[
G^Y_m:= \{(y_1, y_2)\in \mathbbm{R}^2: g(T_1(y_1)/\log m,T_2(y_2)/\log m)=1\},
\]
where $T_i(y)=-\log\{1-\hat{F}_{Y_i}(y)\}$ and $\hat{F}_{Y_i}(y)$ is an estimate of $F_{Y_{i}}(y)$, making use of the probability integral transform. This allows for greater flexibility in the possible marginal distributions, including the possibility of non-equal margins.

\begin{figure}[t]
    \centering
    \includegraphics[width=8cm]{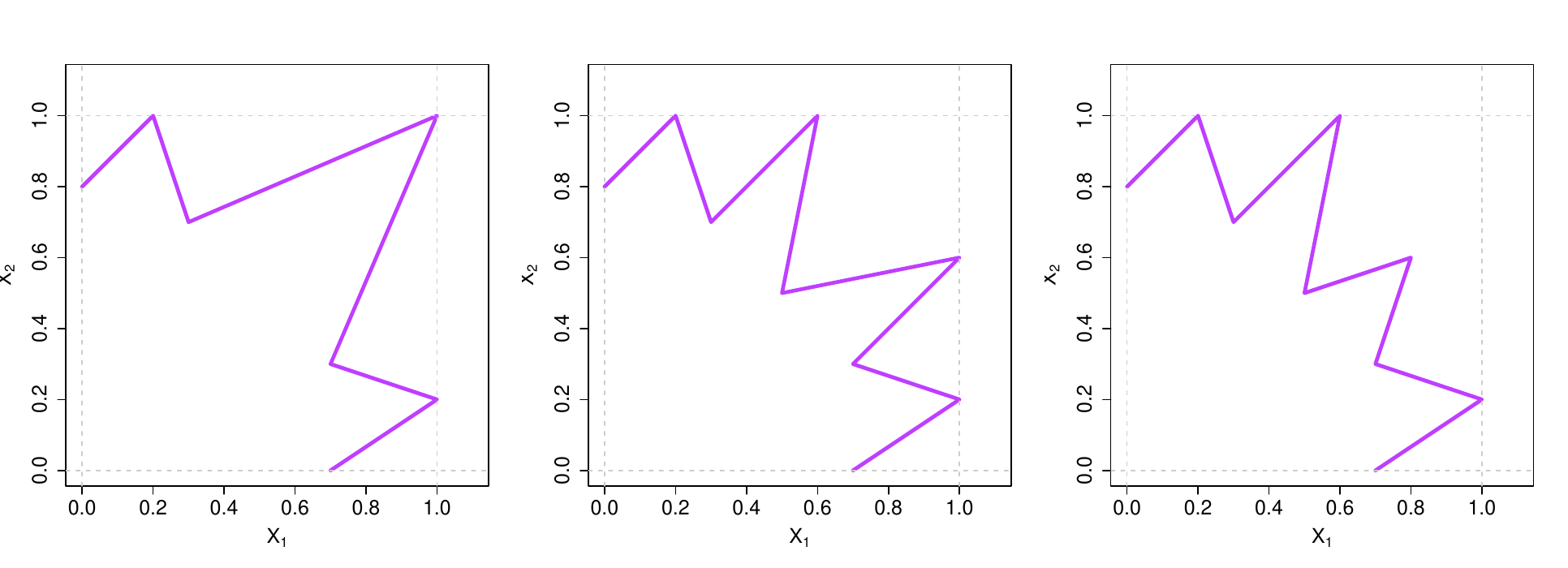}
    \caption{Examples of mixture structures in the set $G$. Left: mixture structure with asymptotic dependence, but the possibility of $X_1$ and $X_2$ both being large while the other is of smaller order. Centre: mixture structure with asymptotic independence, but the largest values of $X_1$ associated with more than one smaller order value of $X_2$ (and vice versa). Right: mixture structure with a sub-asymptotic component.}
\label{fig:mixtureExamples}
\end{figure}

We have discussed the possibility of having mixture structures in the extremal dependence, and there are different ways that this could arise. The framework of \cite{Simpson2020} identifies mixture structures where $X_1$ and $X_2$ can be simultaneously large but with the possibility of having each variable large while the other is of smaller order. Such a structure is indicated by $\eta=\max_{\delta<1}\tau_1(\delta)=\max_{\delta<1}\tau_2(\delta)=1$; examples of sets $G$ with this feature are given in Figure~\ref{fig:mixtureExamples} (left) and the asymmetric logistic copula in Figure~\ref{fig:gaugeEx}. The framework of \cite{Tendijck2021} also allows for a mixture structure in the extremal dependence but with the possibility of $\eta<1$; this arises if $G$ intersects at least one of the lines $x_1=1$ and $x_2=1$ at more than one separate location, as in Figure~\ref{fig:mixtureExamples} (central). Our estimator $\hat G$ can capture such non-convex shapes in the asymmetric logistic case, and therefore has the potential to be further exploited to test for mixture structures. To do this, the scaling/truncation step may need to be altered to ensure that $\hat G$ intersects $x_1=1$ and $x_2=1$ an appropriately many times, perhaps by first testing for convexity of the radial quantile estimates over $w$. A further interesting mixture possibility considered by \cite{Tendijck2021} is that of sub-asymptotic mixture components, as in Figure~\ref{fig:mixtureExamples} (right), where for large $X_1$, e.g., when $x_1=0.7$, there is a mixture of two possible ranges of $X_2$ values. Here, the largest mixture component has a `point' at $(0.8,0.6)$ that does not reach the line $x_1=1$; this should also be taken into account if extending our approach as mentioned above. 

The approach could be adapted to handle higher dimensional problems. An equivalent set $G$ can be defined in higher dimensions; \cite{Nolde2014} presents theoretical results on the coefficient of asymptotic independence in any dimension; \cite{Nolde2021} provides a result equivalent to~\eqref{eqn:NWtau} for calculating the indices of \cite{Simpson2020} in higher dimensions; and in the conditional approach of \cite{Heffernan2004}, pairwise results can determine the relevant normalising functions. Our approach involves transforming to pseudo-polar coordinates, with only a one-dimensional angular component in the bivariate case. In a $d$-dimensional setting, the angular component would have dimension $d-1$; this raises questions about how to select the angles used for local estimation to ensure reliability without a high computational cost. Moreover, the selection of the spline functions for $\hat G$ comes with added complexity in higher dimensions, as different choices may be needed for different subsets of the variables. Future work that extends our approach to higher dimensions, should involve developing a strategy for selecting estimation points and spline functions in order to achieve scalable but reliable inference. 

A further issue related to higher dimensional problems involves ensuring consistency across different dimensions. Suppose we have variables $\bm{X}=(X_i:i\in\mathcal{D})$, for $\mathcal{D}=\{1,\dots,d\}$. Using results on projections of sample clouds, \cite{Nolde2021} makes links between gauge functions of lower dimensional subsets (for instance, variables indexed by $\mathcal{C}\subset\mathcal{D}$) to the gauge function associated with the full vector $\bm{X}$. If our method were to be extended to higher dimensions, estimates of extremal dependence features for the variables $\bm{X}_\mathcal{C}=(X_i:i\in\mathcal{C})$ could be obtained by estimating the asymptotic boundary of the scaled sample cloud for any set of variables indexed by $\mathcal{C}^*$ with $\mathcal{C}\subseteq\mathcal{C}^*\subseteq\mathcal{D}$. Ideally, we would ensure consistent estimation from these different boundary estimates.

\paragraph{Code:} Code to run the methods presented in this paper is available online in the GitHub repository:\\ \url{https://github.com/essimpson/self-consistent-inference}.

\bibliographystyle{apalike}
\bibliography{refs}

\newpage

\renewcommand{\thesection}{\Alph{section}}
\renewcommand{\thefigure}{\Alph{figure}}
\setcounter{figure}{0} 
\setcounter{section}{0} 

\begin{center}
{\hrule\vspace{10pt} \LARGE  Supplementary Material for `Estimating the limiting shape of bivariate scaled sample clouds: with additional benefits of self-consistent inference for existing extremal dependence properties'\\}
\vspace{10pt}\author{Emma S.\ Simpson$^1$ and Jonathan A.\ Tawn$^2$\\ 
\normalsize{$^1$Department of Statistical Science, University College London, WC1E 6BT, U.K.}\\
\normalsize{$^2$School of Mathematical Sciences, Lancaster University, LA1 4YF, U.K.}}
\vspace{10pt}\hrule
\end{center}
\vspace{10pt}
\begin{abstract}
The aim of this Supplementary Material is to provide additional details for the paper `Estimating the limiting shape of bivariate scaled sample clouds: with additional benefits of self-consistent inference for existing extremal dependence properties'. In Section~\ref{SM:Glinks}, we provide an illustration of the theoretical links between the set $G$ and the extremal dependence features under consideration. In Section~\ref{subsec:tuning}, we discuss the selection of tuning parameters for our limit set estimates $\hat G^L$ and $\hat G$, while Section~\ref{SM:estimationIllustration} provides an illustration of our scaling/truncation approach and how we obtain estimates of several bivariate extremal dependence features from $\hat G$. In Section~\ref{SM:existingEstimators}, we provide additional details on the existing estimators for the extremal dependence features we consider. Most of the remaining sections provide additional simulation results for our approach: in Section~\ref{SM:qSensitivity}, we present estimates of $\hat G$ to test sensitivity to the extrapolation quantile $q$; in Section~\ref{SM:tauResults}, we give results on estimating $\tau_1(\delta)$ for additional copula dependence parameters; Section~\ref{SM:lambdaResults} covers estimation of $\lambda(\omega)$; Section~\ref{SM:HTResults} provides results on the $(\alpha_i,\beta_i)$ parameters of the conditional extremes approach; Section~\ref{SM:alog} provides additional simulation results for the asymmetric logistic copula; in Section~\ref{SM:size100000}, we demonstrate the performance of our approach on samples of size $n=100,000$; and in Section~\ref{SM:bootstrap}, we test whether bootstrapping is a reasonable approach to uncertainty assessment for our estimator. A summary of our full approach to estimate $G$ is provided in Section~\ref{SM:algorithm}.
\end{abstract}

\section{Illustration of the link between extremal dependence features and the set $G$}\label{SM:Glinks}
In Section~\ref{sec:NWtheory} of the main paper, we discussed results from \cite{Nolde2014} and \cite{Nolde2021}, linking the set $G$ to various bivariate extremal dependence features. Now, we provide further discussion and a pictorial demonstration of the theoretical results.

\begin{figure}[t]
    \centering
    \includegraphics[width=0.75\textwidth]{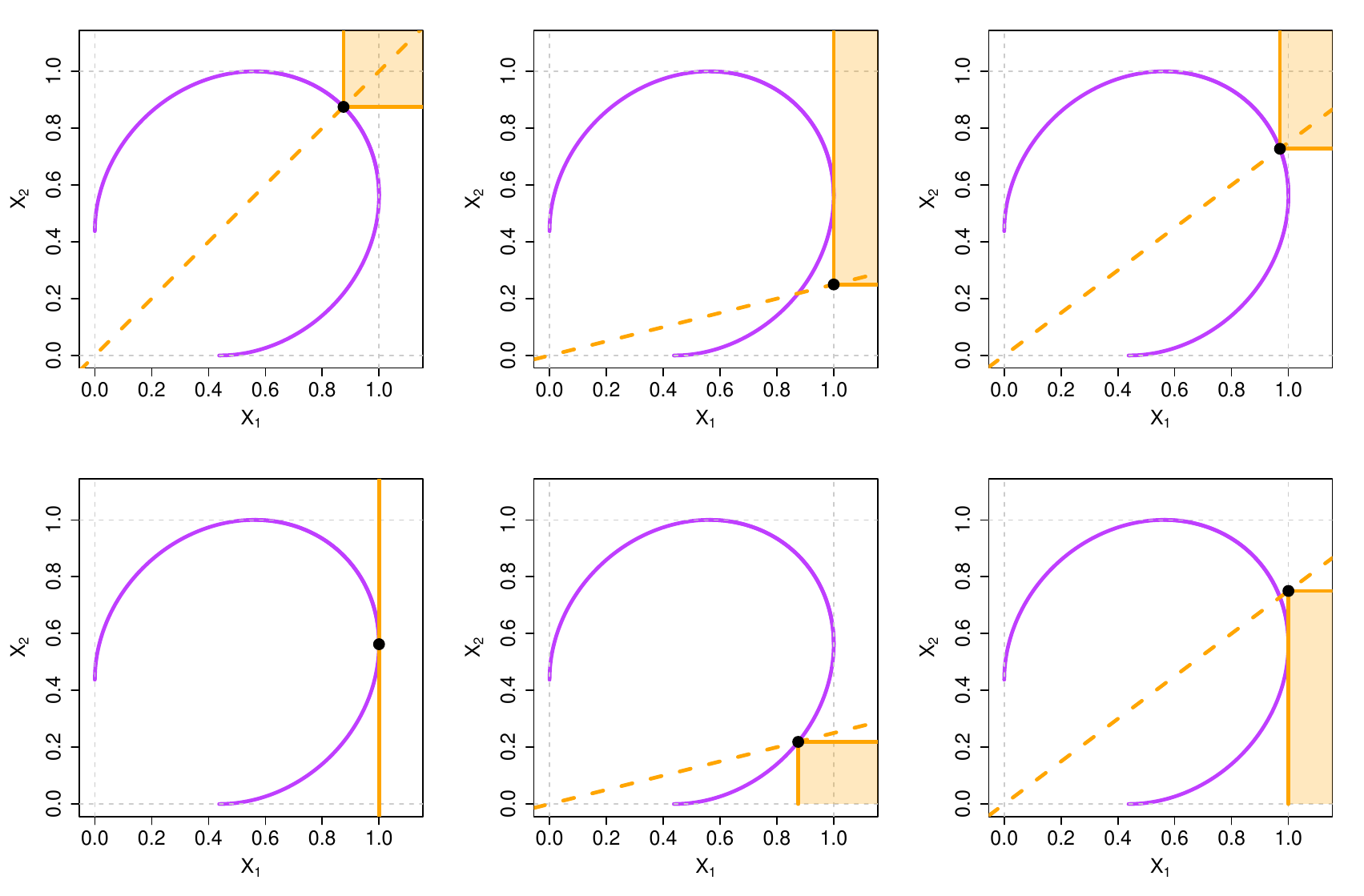}
    \vspace{-0.75cm}
    \caption{Examples of the set $G$ (purple) and demonstration of calculating $\eta$, $\lambda(4/5)$, $\lambda(4/7)$, $\tau_1(3/4)$, $\tau_1(1/4)$ and $\alpha_1$ (clockwise from top left) for a Gaussian model with correlation parameter $\rho=0.75$. 
    Top left: the set $[\eta,\infty]^2$ (orange region), and the point $(\eta,\eta)$ (black circle) which lies on $x_2=x_1$ (orange dashed line). Top, centre and right: the sets $s_\omega S_\omega$ (orange regions) for $\omega=4/5,4/7$, and the points $\left(s_\omega,\frac{1-\omega}{\omega}s_\omega\right)$ (black), lying on $x_2=\frac{1-\omega}{\omega} x_1$ (orange dashed line). Bottom, right and centre: the sets $\tau_1(\delta)S_{1,\delta}$ (orange regions), for $\delta=3/4,1/4$, and the points $\left(\tau_1(\delta),\delta\tau_1(\delta)\right)$ (black), lying on $x_2=\delta x_1$ (orange dashed line). Bottom left: the line $x_1=1$ (orange) and the point $(1,\alpha_1)$ (black).}
    \label{fig:featureEx}
\end{figure}

First, \cite{Nolde2014} links the coefficient $\eta$ from \cite{Ledford1996} using result~\eqref{eqn:Neta} in the main paper. The result can be thought of as moving the set $[1,\infty]^2$ towards the origin, along the line $x_2=x_1$, until it just intersects the set $G$. We demonstrate this feature in the top left panel of Figure~\ref{fig:featureEx} for a Gaussian copula with $\rho=0.75$. In this case, $\eta=(1+\rho)/2=0.875$, corresponding to the known value of $\eta$ for this model. We note that although the intersection in this example happens when $x_1=x_2=\eta$, this does not have to be the case, and it can occur on one of the edges of the orange rectangle with $x_1\neq x_2$.

The central and right panels of the top row of Figure~\ref{fig:featureEx} demonstrate the sets $s_\omega S_\omega$, used to calculate $\lambda(\omega)$ from \cite{Wadsworth2013} through equation~\eqref{eqn:NWlambda}. These plots are again for the Gaussian copula with $\rho=0.75$ and with $\omega=4/5,4/7$. This example highlights the need to consider the whole set $S_\omega$ when deducing $\lambda(\omega)$, not just the vertex $x_2=\frac{1-\omega}{\omega}x_1$. When $\omega=4/5$, the intersection of $G$ and $s_\omega S_\omega$ occurs on the edge of the rectangle, with $s_\omega=1$ and $\lambda(\omega)=\max(\omega,1-\omega)=4/5$. In contrast, for $\omega=4/7$, the intersection does occur on the line $x_2=\frac{1-\omega}{\omega}x_1$, with the value of $\lambda(4/7)$ given by the first Gaussian $\lambda(\omega)$ case in Table~\ref{tab:parameterValues} of the main paper.

In a similar way to the $\eta$ representation of \cite{Nolde2014}, we can think of $\tau_1(\delta)$ as taking the set $S_{1,\delta}$ and moving it along the line $x_2=\delta x_1$ towards the origin until it intersects the set $G$ (where, again, the intersection does not necessarily occur on the line $x_2=\delta x_1$). This is demonstrated in the right and central panels of the bottom row of Figure~\ref{fig:featureEx} for the same Gaussian model, taking $\delta=0.75,0.25$, respectively. We have $\tau_1(0.75)=1$, and $\tau_1(0.25)<1$, with its exact form given in Table~\ref{tab:parameterValues}. 

Finally, we demonstrate the calculation of $\alpha_1$ for our Gaussian example in the bottom left panel of Figure~\ref{fig:featureEx}. Here, $\alpha_1=\rho^2=0.75^2$.
 
\section{Tuning parameter selection}\label{subsec:tuning}
Our proposed estimation procedure for the set $G$, presented in Section~\ref{sec:method} of the main paper, involves the choice of several tuning parameters. For $\hat G^L$ in Section~\ref{subsec:Gestimation2}, the first step is to choose the value of $k$ and the angular values $w_j^*$, for $j\in J_k$. The discussion in Section~\ref{subsec:limitSet} of the main paper highlighted that whether or not the point $(1,1)$ belongs to the set $G$ can play a crucial role in distinguishing between asymptotic dependence and asymptotic independence. As a result, the value of $G$ at $w=1/2$ is of importance, so 1/2 should be taken to be one of the estimation angles $w_j^*$. We only take $w_j^*$ values within the range of observed angles, i.e., setting $w^m=\min_{i=1,\dots,n}w_i$ and $w^M=\max_{i=1,\dots,n}w_i$, we have $w_j^*\in[w^m,w^M]$ for all $j\in J_k$. Natural options for choosing the $w_j^*$ values include splitting the interval $[w^m,w^M]$ into equal segments or using empirical quantiles of $\{w_1,\dots,w_n\}$. The former has the drawback that several $w_j^*$ values could be selected without any nearby values in $\{w_1,\dots,w_n\}$. In such cases, the neighbourhood in~\eqref{eqn:Rneighbours} would need to be relatively large to allow for inference, but this would come at the cost of obtaining reliable estimates of $r_q(w)$, as that may change considerably over this range of $w$. We therefore prefer the latter approach, with $w_j^*$ as the $(j-1)/(k-1)$th empirical quantile of $\{w_1,\dots,w_n\}$, for $j=1,\dots,k-1$, and $w_k^*=1/2$. We take $k$ to be odd so that if the angular distribution is symmetric, we can have an equal number of $w_j^*$ values either side of 1/2. Our studies suggest that the choice of $k$ is not particularly crucial to the performance of the estimator, and we set $k=199$ as a default.

\begin{figure}[t]
    \centering
    \includegraphics[width=0.7\textwidth]{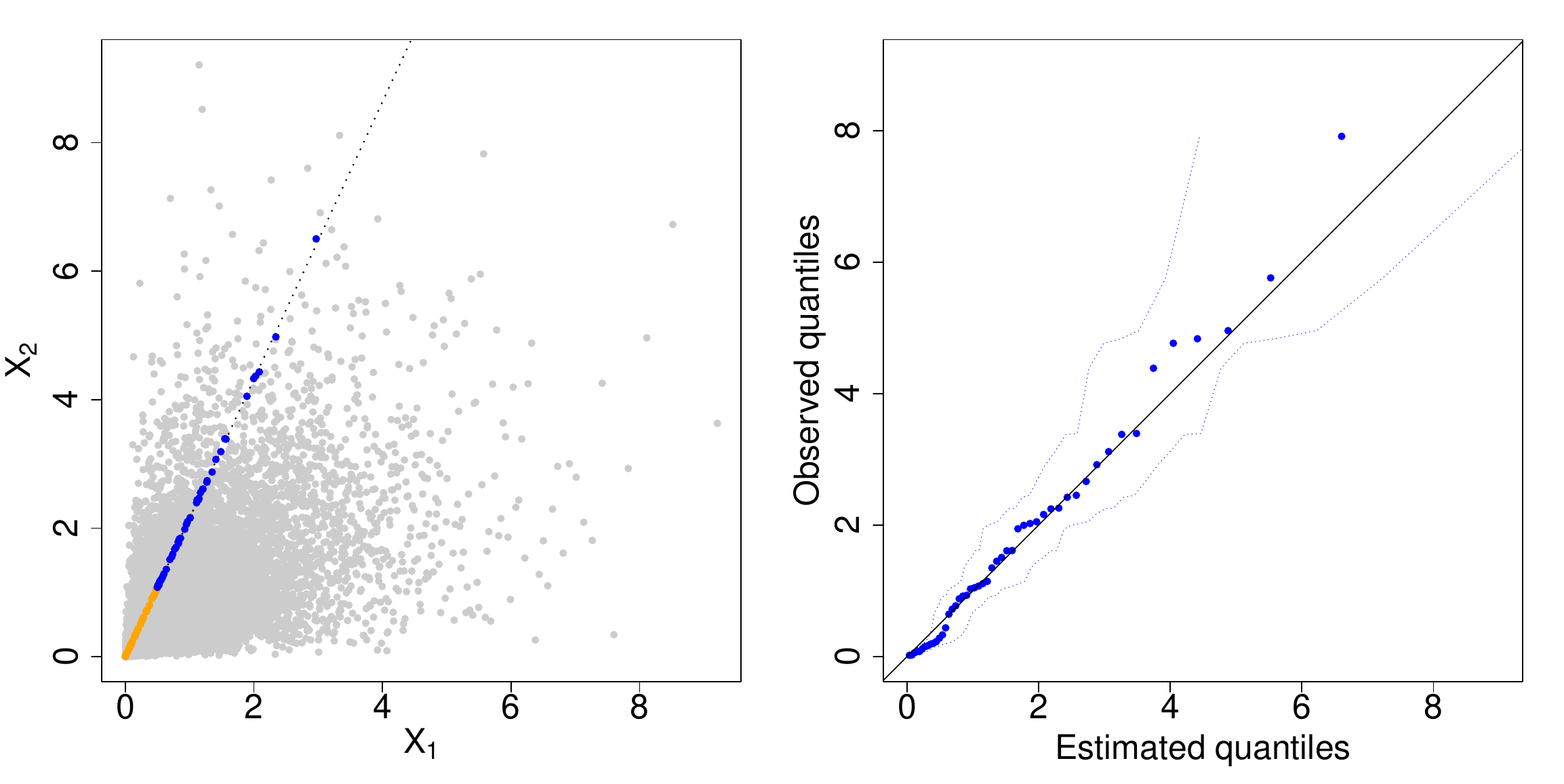}
    \vspace{-0.5cm}
    \caption{Left: a sample of size 10,000 from an inverted logistic distribution with $\gamma=0.5$ (grey) and an example of a set $\mathcal{R}_{w_j^*}$, with $|\mathcal{R}_{w_j^*}|=m=100$ (orange: points in $\mathcal{R}_{w_j^*}$ below the empirical 0.5 quantile; blue: points in $\mathcal{R}_{w_j^*}$ above the empirical 0.5 quantile). Right: QQ-plot for the fitted GPD model for exceedances of $\mathcal{R}_{w_j^*}$ above the 0.5 quantile. The blue dashed lines represent approximate 95\% confidence intervals for the pointwise empirical quantiles.}
    \label{fig:Gexample}
\end{figure}

Next consider the sizes of each radial neighbourhood $\epsilon_{w_j^*}~(j\in J_k)$. We allow these to depend on the angle $w_j^*$, in order to control the size of each set $\mathcal{R}_{w_j^*}$. In particular, we fix each $\epsilon_{w_j^*}$ such that $|\mathcal{R}_{w_j^*}|=m$, for some choice of $m$. That is, $\mathcal{R}_{w_j^*}$ corresponds to the radial values of the $m$ nearest angular neighbours of $w_j^*$. The GPD threshold $u_{w_j^*}$, or equivalently the value $q_u\in[0,q]$, also needs selecting. Figure~\ref{fig:Gexample} demonstrates an example of a single set $\mathcal{R}_{w_j^*}$ for the same data as in Figure~\ref{fig:Gexample2}. The QQ-plot in Figure~\ref{fig:Gexample} (right) suggests that taking $u_{w_j^*}$ as the empirical median of $\mathcal{R}_{w_j^*}$ (i.e., $q_u=0.5$) is a reasonable choice, and we have observed the same behaviour over a range of different examples during our investigations. For simplicity, we therefore fix $q_u=0.5$ in the paper. The final tuning parameter to select for the local approach is the value of $q$ in~\eqref{eqn:radialQuantile}. We have found values of $q$ between 0.9 and 0.9999 to work well in practice, and take $q=0.999$ subsequently. Sensitivity to this parameter is investigated in Section~\ref{SM:qSensitivity}, where the method appears robust to a reasonable range of choices of $q$.

When using the GAM framework to produce the smoothed estimates proposed in Section~\ref{subsec:GestimationSmooth} of the main paper, we must select the number of knots ($\kappa$) and degree of the corresponding spline functions. For the reasons we discussed earlier, $w=1/2$ is taken as one of the knots, which also suggests that an odd value of $\kappa$ is reasonable. In practice, we suggest spacing the $\kappa$ interior spline knots evenly throughout $[w^m,w^M]$, then adjusting the central one to be exactly 1/2. Any required exterior knots are fixed at 0 and 1. We have found $\kappa=7$ to be a reasonable choice in practice, illustrating limited sensitivity to this tuning parameter in Section~\ref{SM:alog}. 

\section{Illustrations of our scaling/truncation and parameter estimation approaches}\label{SM:estimationIllustration} 
\begin{figure}[!htbp]
    \centering
    \includegraphics[width=\textwidth]{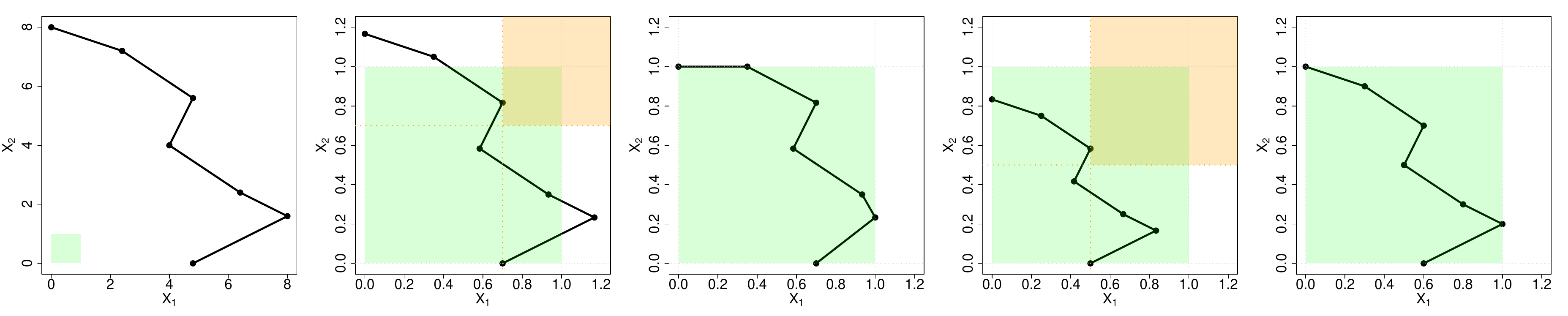}
    \caption{An illustration of our scaling approach. From left to right, the first plot shows an example set of pre-scaled points $(\tilde{x}_{1,j},\tilde{x}_{2,j})$, $j\in J_k$ (black dots), with $k=7$ for illustrative purposes. The second plot demonstrates step 1 with $\hat\eta_H=0.7$ including the set $[0.7,\infty]^2$ (orange region) and $\hat G^L_\eta$ (black points) and the third plot shows the corresponding final set $\hat G^L$ with points outside $[0,1]^2$ transformed onto the boundary (step 2a). The fourth plot is equivalent to the second but with $\hat\eta_H=0.5$, and the final plot shows the corresponding final set $\hat G^L$ with the final transformation ensuring intersection with $x_1=1$ and $x_2=1$ (step 2b). The green region in each plot shows the set $[0,1]^2$.}
    \label{fig:scalingDemo}
\end{figure}

\begin{figure}[!htbp]
    \centering
    \includegraphics[width=0.9\textwidth]{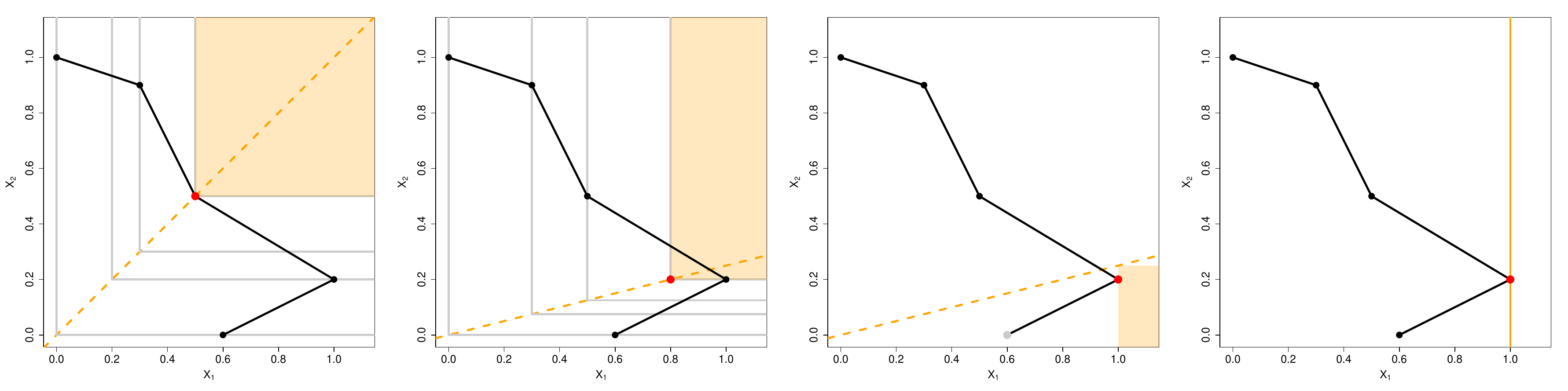}
    \caption{A demonstration of how to obtain estimates of $\eta$, $s_{0.8}$ (for subsequent estimation of $\lambda(0.8)$), $\tau_1(0.25)$ and $\alpha_1$ from the estimate of $G$. Here, $\hat G$ uses $k = 5$ for illustrative purposes, but in practice $k$ should be much larger. The black points in each plot demonstrate an estimated set $\hat G$. Left: the boundaries of the candidate sets $[s^*_j,\infty]^2$ for each point in $\hat G$ (grey); the set $[\hat\eta_G,\infty]^2$ (orange region); and the point $(\hat\eta_G,\hat\eta_G)$ (red) which satisfies $x_1=x_2$ (orange dashed line). Left-centre: the line $x_2=\frac{1-0.8}{0.8}x_1=0.25x_1$ (orange dashed line); the boundaries of the candidate sets $[s^*_{0.8,j},\infty]\times[0.25s^*_{0.8,j},\infty]$ for each point in $\hat G$ (grey); the set $[\hat s_{0.8},\infty]\times[0.25\hat s_{0.8},\infty]$ (orange region); the point of intersection of the orange region with $\hat G$ (red). Right-centre: the line $x_2=0.25 x_1$ (orange dashed line); the set $(\tau_1(0.25),\infty]\times[0,0.25\tau_1(0.25)]$ (orange region); the point of intersection of the orange region with $\hat G$ (red) and the other candidate for this intersection point (grey). Right: the line $x_1=1$ (orange) and its intersection with the set $\hat G$ (red).}
    \label{fig:estimateExamples}
\end{figure}

The scaling/truncation approach discussed in Section~\ref{subsec:Gestimation2} of the main paper is demonstrated in Figure~\ref{fig:scalingDemo}. This allows for estimation of a finite set of points belonging to the set $G\subset[0,1]^2$, with our final estimate given by $\hat G^L=\left\{(\hat{x}^G_{1,j},\hat{x}^G_{2,j}):j\in J_k\right\}$. In Figure~\ref{fig:estimateExamples}, we provide an illustration of our estimation procedure for $\eta$, $s_\omega$ (for $\lambda(\omega)$), $\tau_1(\delta)$ and $\alpha_1$. This accompanies the discussion in Section~\ref{subsec:parameterEstimation} of the main paper.

\vspace{0.5cm}

\section{Existing estimation techniques}\label{SM:existingEstimators}
In this section, we present some existing estimators for the extremal dependence properties in Section~\ref{subsec:BEVfeatures} of the main paper. We use these for comparison to our proposed estimators in Section~\ref{sec:results}. \cite{Ledford1996} estimate $\eta$ using a maximum-likelihood based approach, which results in a Hill-type estimator \citep{Hill1975}, but on exponential margins. Setting $M=\min(X_1,X_2)$, and assuming we have $n_{u,H}$ observations of $M$ above some high threshold $u_H$, denoted $m^*_1,\dots,m^*_{n_{u,H}}$, this estimator is 
\begin{align*}
\tilde\eta_{H} = \frac{1}{n_{u,H}}\sum\limits_{i=1}^{n_{u,H}} \left(m^*_i-u_H\right).
\end{align*}
Alternative estimation procedures for $\eta$ include the non-parametric approaches of \cite{Peng1999} and \cite{Draisma2004}. These aim to avoid inconsistencies that arise in carrying out estimation using different marginal distributions, which can result in issues when quantifying uncertainty. Suppose we have $n$ pairs of observations $(x_{1,1},x_{1,2}),\dots,(x_{n,1},x_{n,2})$, and denote by $x^{(j)}_i$ the $j$th largest value of component $i\in\{1,2\}$, i.e., $x_i^{(n)}\leq x_i^{(n-1)}\leq\dots\leq x_i^{(1)}$. In both approaches, for each $j\in\{1,\dots,n\}$, the first step is to define the quantity $s_n(j) = \sum_{\ell=1}^n \bm{1}\left\{x_{\ell,1} \geq x^{(j)}_1, x_{\ell,2} \geq x^{(j)}_2\right\}$, with $\bm{1}(E)$ the indicator of event $E$. The estimators $\tilde\eta_P$ of \cite{Peng1999} and $\tilde\eta_D$ of \cite{Draisma2004} are
\begin{align*}
    \tilde\eta_{P} = \frac{\log 2}{\log\left\{s_n(2c)\right\}-\log\left\{s_n(c)\right\}}, \qquad
    \tilde\eta_D = \frac{\sum_{j=1}^cs_n(j)}{cs_n(c)- \sum_{j=1}^cs_n(j)},
\end{align*}
where $c$ relates to the number of exceedances above a high threshold. Further details on these estimators can be found in \cite{Beirlant2004}. Although $\eta\in(0,1]$, all three of these estimators can result in values greater than one, so the truncated estimators $\hat\eta_H=\min(\tilde\eta_H,1)$, $\hat\eta_P=\min(\tilde\eta_P,1)$ and $\hat\eta_D=\min(\tilde\eta_D,1)$ are preferable. Our estimator from Section~\ref{subsec:parameterEstimation} of the main paper guarantees that $\hat\eta_G\in(0,1]$ due to the scaling procedure proposed in Section~\ref{subsec:Gestimation2}.

\cite{Wadsworth2013} also discuss the use of a Hill-type estimator for $\lambda(\omega)$, analogous to $\hat\eta_H$, but taking $M_\omega=\min\left\{X_1/\omega,X_2/(1-\omega)\right\}$, for $\omega\in[0,1]$. If we have $n_{u,\omega}$ observations of $M_\omega$ above a high threshold $u_\omega$, denoted $m^*_{\omega,1},\dots,m^*_{\omega,n_{u,\omega}}$, the estimator is
\begin{align*}
\tilde\lambda_{H}(\omega) = \left\{\frac{1}{n_{u,\omega}}\sum\limits_{i=1}^{n_{u,\omega}} \left(m^*_{\omega,i}-u_\omega\right)\right\}^{-1},
\end{align*}
i.e., the reciprocal of the estimated mean excess of $M_\omega$ above the threshold $u_\omega$. Again, the truncated estimator $\hat\lambda_H(\omega)=\min\left(\tilde\lambda_H(\omega),1\right)$ is preferred. Similarly, \cite{Simpson2020} propose a Hill-type estimator for the coefficients $\tau_1(\delta)$, $\delta\in(0,1]$. The idea is to slightly alter representation~\eqref{eqn:tauDef}, and instead consider $\Pr(X_1>x,X_2\leq \delta X_1),$ for which estimation is simpler; \cite{Nolde2021} prove this has index of regular variation $\tau_1(\delta)$. Let $M_\delta = \{X_1>0 : X_2 \leq \delta X_1\}$. Suppose we have $n_\delta\leq n$ observations of $X_1$ where $X_2 \leq \delta X_1$, and that $n_{u,\delta}$ of these, denoted $m^*_{\delta,1},\dots, m^*_{\delta,n_{u,\delta}}$, are above some high threshold $u_\delta$. Then, 
\begin{align*}
\tilde\tau_{H,1}(\delta) = \frac{1}{n_{u,\delta}}\sum\limits_{i=1}^{n_{u,\delta}} \left(m^*_{\delta,i}-u_\delta\right).
\end{align*}
Again, we could have $\tilde\tau_{H,1}(\delta)>1$ here, so we truncate to give $\hat\tau_{H,1}(\delta)=\min\{\tilde\tau_{H,1}(\delta),1\}$.

\section{Sensitivity to $q$}\label{SM:qSensitivity}
\begin{figure}[!htbp]
    \centering
    \includegraphics[width=0.9\textwidth]{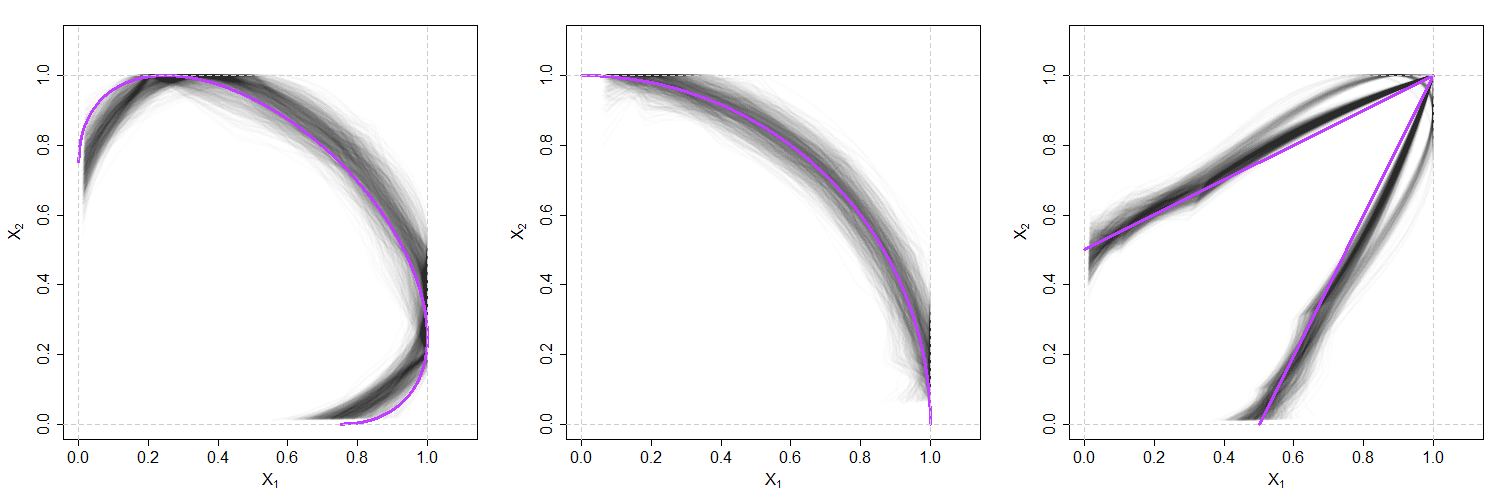}
    \includegraphics[width=0.9\textwidth]{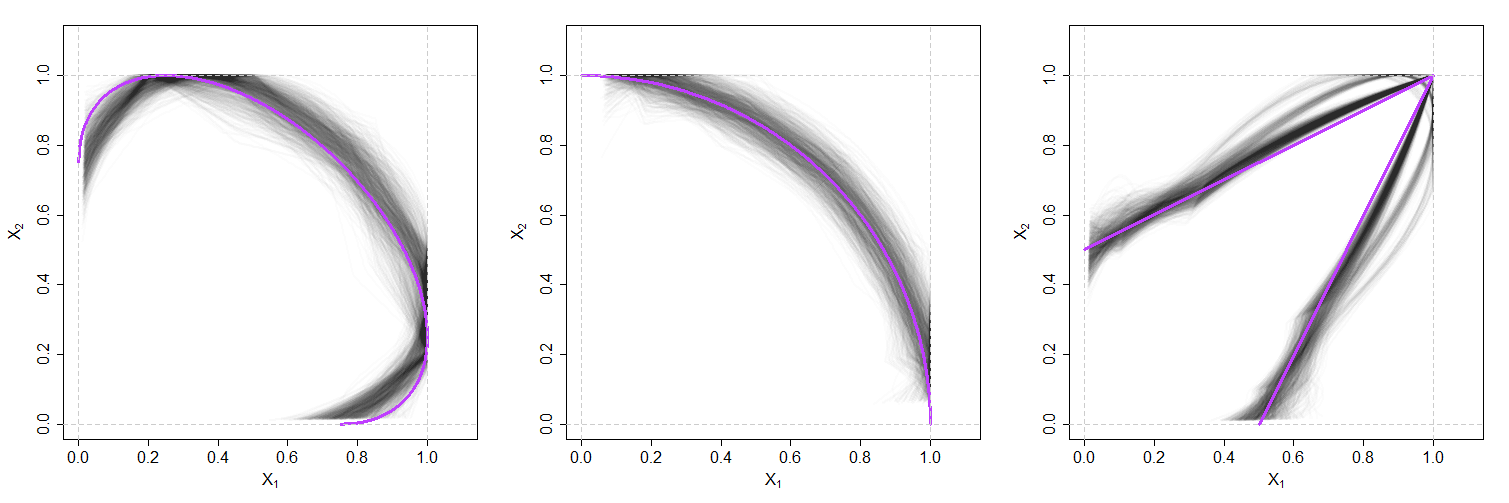}
    \caption{Estimates of the sets $G$ for the Gaussian (left), inverted logistic (centre) and logistic (right) copulas, with the corresponding $\rho$ or $\gamma$ parameters set to 0.5. The sample size in each case is $n=10,000$ and 1000 estimates are shown for each model (grey). The true sets $G$ are shown in purple. The radial extrapolation quantile is set as $q=0.99$ (top row) and $q=0.9999$ (bottom row).}
    \label{fig:Gestimates99}
    \vspace{1cm}
    \includegraphics[width=0.9\textwidth]{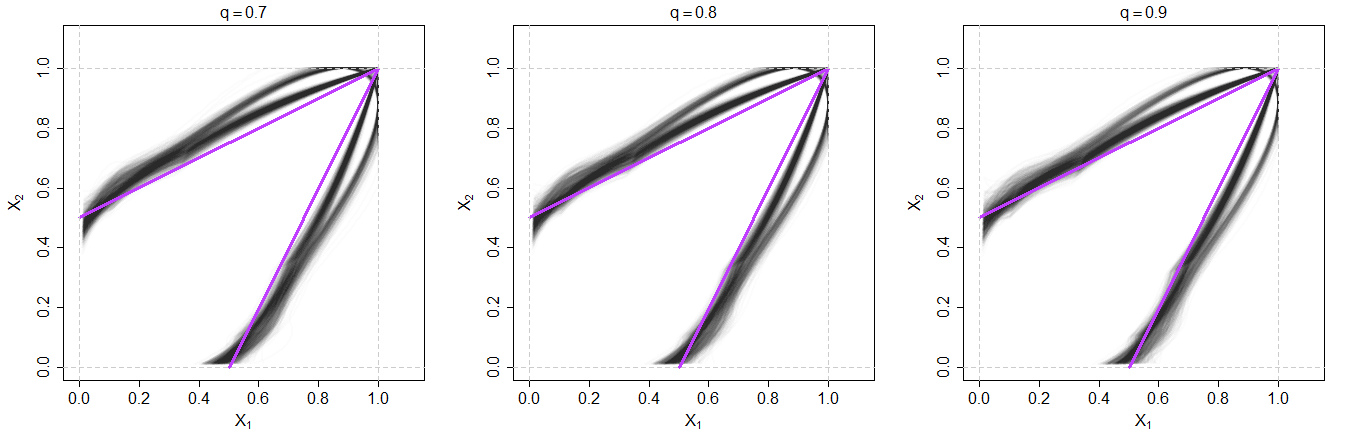}
    \caption{Estimates of the sets $G$ for the logistic copula with $\gamma=0.5$. The sample size in each case is $n=10,000$ and 1000 estimates are shown for each model (grey). The true sets $G$ are shown in purple. The radial extrapolation quantile is set as $q=0.7$ (left), $q=0.8$ (centre) and $q=0.9$ (right).}
    \label{fig:Gestimates_lowerq}
\end{figure}

Our method involves the selection of several tuning parameters. Intuitively, one of the most important is the value of $q$, representing the quantile of interest for the radial distributions before scaling onto $[0,1]^2$. For all simulations and applications presented in the paper, we set $q=0.999$, but we now present estimates of the set $G$ for alternative quantile choices. All other tuning parameters are fixed as in Section~\ref{subsec:simulationStudy} of the main paper. As in Figure~\ref{fig:Gexample3} of the paper, we present results for datasets of size 10,000 simulated from Gaussian, inverted logistic and logistic models, repeating each test 1000 times. The results are shown in Figure~\ref{fig:Gestimates99} for $q=0.99$ and $q=0.9999$. Here, we choose only values of $q$ close to one in order to appropriately estimate the \emph{limiting} shape of the scaled sample cloud; taking $q$ to be smaller than this risks estimating a sub-asymptotic level set that is different in shape to the limit, which we explore later in this section. Overall, there is very little disparity between the estimates with $q$ set as 0.99, 0.999 or 0.9999, since differences in the $r_q(w)$ estimates are offset by the subsequent scaling/truncation steps. There is some evidence that the estimates for the logistic model deteriorate as we increase $q$, with the least desirable quadratic spline option being chosen slightly more often for $q=0.9999$ than $q=0.99$, although the linear splines are still chosen in around 78\% of cases. Hence, the estimates are overall not particularly sensitive to the choice of $q$, suggesting setting $q=0.999$ is a reasonable choice.

To highlight the need to take $q$ close to one, we now consider the logistic case in further detail. In Figure~\ref{fig:Gestimates_lowerq}, we present results of a similar simulation study, but where the radial quantile level is set to $q=0.7,~0.8$ or $0.9$. The percentage of times that the most desirable, linear splines are selected in these cases is 65\%, 68\% and 76\%, respectively, so we are able to correctly identify asymptotic dependence more often as we increase the value of $q$. This happens since for this copula, the pointed shape we see under asymptotic dependence is a limiting feature, and true level sets at lower quantiles have a smoother, curved shape. Setting $q$ close to one is therefore necessary to improve the estimation of the limiting shape of the scaled sample cloud, $G$.

\section{Additional estimates of $\tau_1(\delta)$}\label{SM:tauResults}
\begin{figure}[!htbp]
    \centering
    \includegraphics[width=0.95\textwidth]{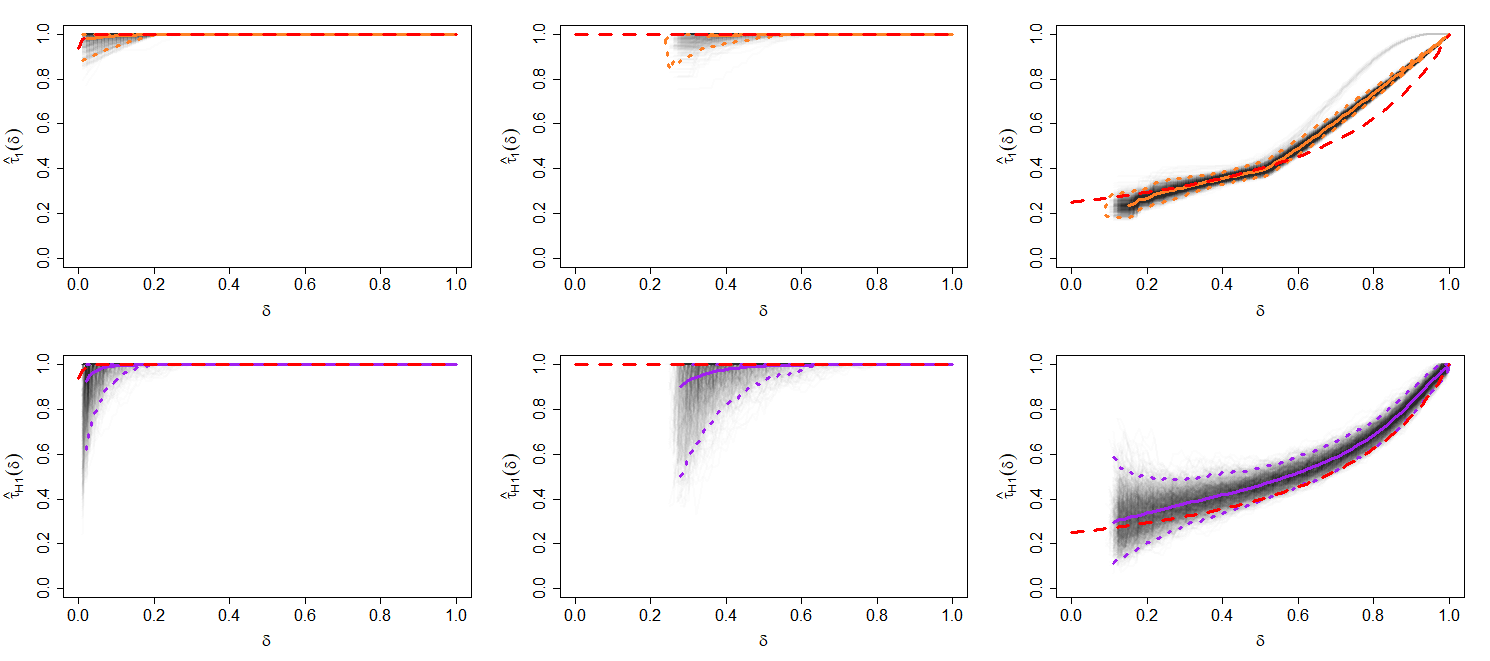}
        \caption{Estimates of $\tau_1(\delta)$, $\delta\in[0,1]$, for data simulated from Gaussian (left), inverted logistic (centre) and logistic (right) models with their corresponding parameters set as 0.25. Each grey line represents an estimate obtained using $\hat\tau_{G,1}(\delta)$ (top) or $\hat\tau_{H,1}(\delta)$ (bottom) over 1000 simulations. The solid and lower/upper dotted lines show the pointwise bootstrapped means and 0.025 and 0.975 quantiles, respectively, for $\hat\tau_{G,1}(\delta)$ (orange) and $\hat\tau_{H,1}(\delta)$ (purple). The true $\tau_1(\delta)$ values are shown in red.}
    \label{fig:tauEstimates0.25}
\end{figure}

\begin{figure}[!htbp]
    \centering
    \includegraphics[width=0.95\textwidth]{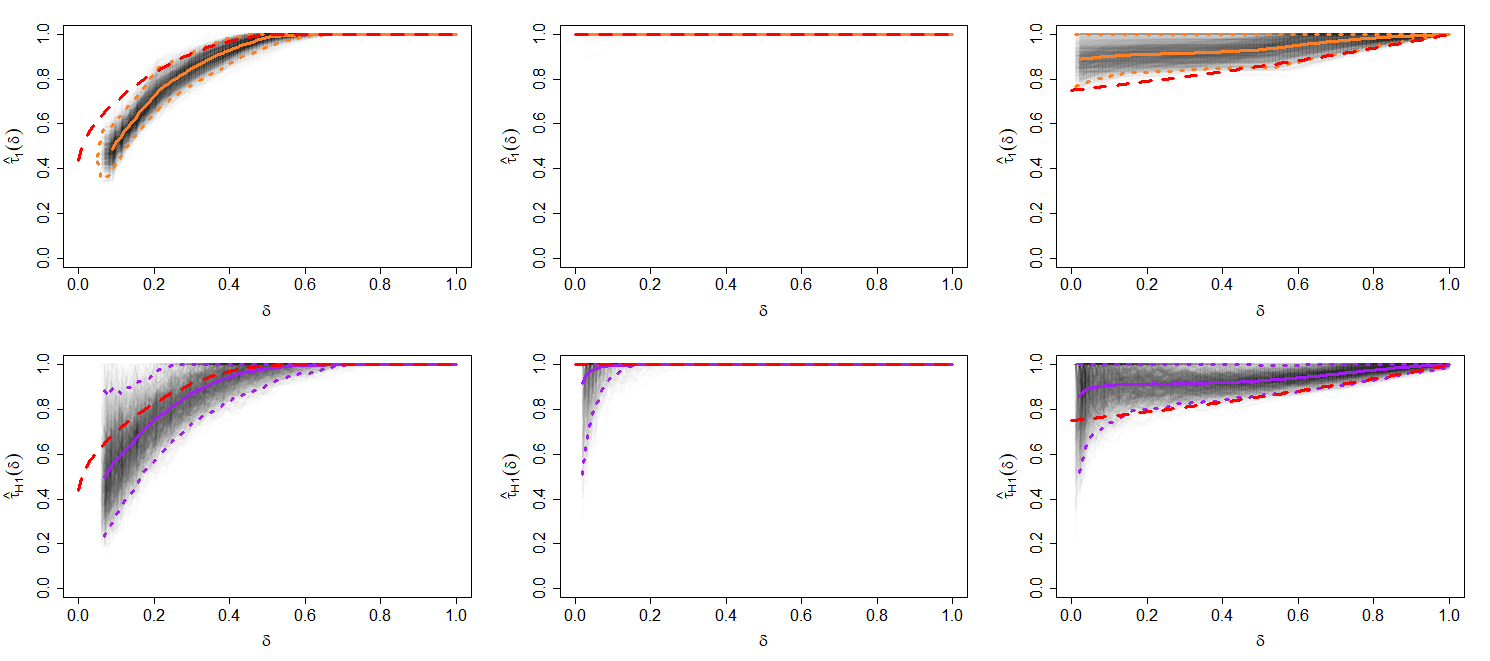}
        \caption{Plots equivalent to Figure~\ref{fig:tauEstimates0.25}, but with the copula parameters set to 0.75.}
    \label{fig:tauEstimates0.75}
\end{figure}

In Figure~\ref{fig:tauEstimates0.5} of the main paper, we presented results on estimation of the parameter $\tau_1(\delta)$ \citep{Simpson2020} for Gaussian, inverted logistic and logistic models with parameters $\rho$ or $\gamma$ set to 0.5. In Figures~\ref{fig:tauEstimates0.25}~and~\ref{fig:tauEstimates0.75}, respectively, we now present equivalent results for $\rho,\gamma=0.25$ and $\rho,\gamma=0.75$, comparing our new estimator to the Hill-type estimator $\hat\tau_{H,1}(\delta)$. We observe similar behaviour to that in the results of the paper. In particular, for smaller values of $\delta$, our $\hat\tau_{G,1}(\delta)$ provides improvements in variance over the $\hat\tau_{H,1}(\delta)$ estimates. In terms of bias, our results are most successful in the inverted logistic cases and for smaller $\rho$ and $\gamma$ parameters. For $\rho=0.75$, we tend to underestimate $\tau_1(\delta)$ for $\delta<0.5$ in the Gaussian case, which is not a problem for the Hill-type estimator, while for $\gamma=0.75$ both approaches yield overestimation in the logistic setting.

\section{Estimates of $\lambda(\omega)$}\label{SM:lambdaResults}

\begin{figure}[!htbp]
    \centering
    \includegraphics[width=0.95\textwidth]{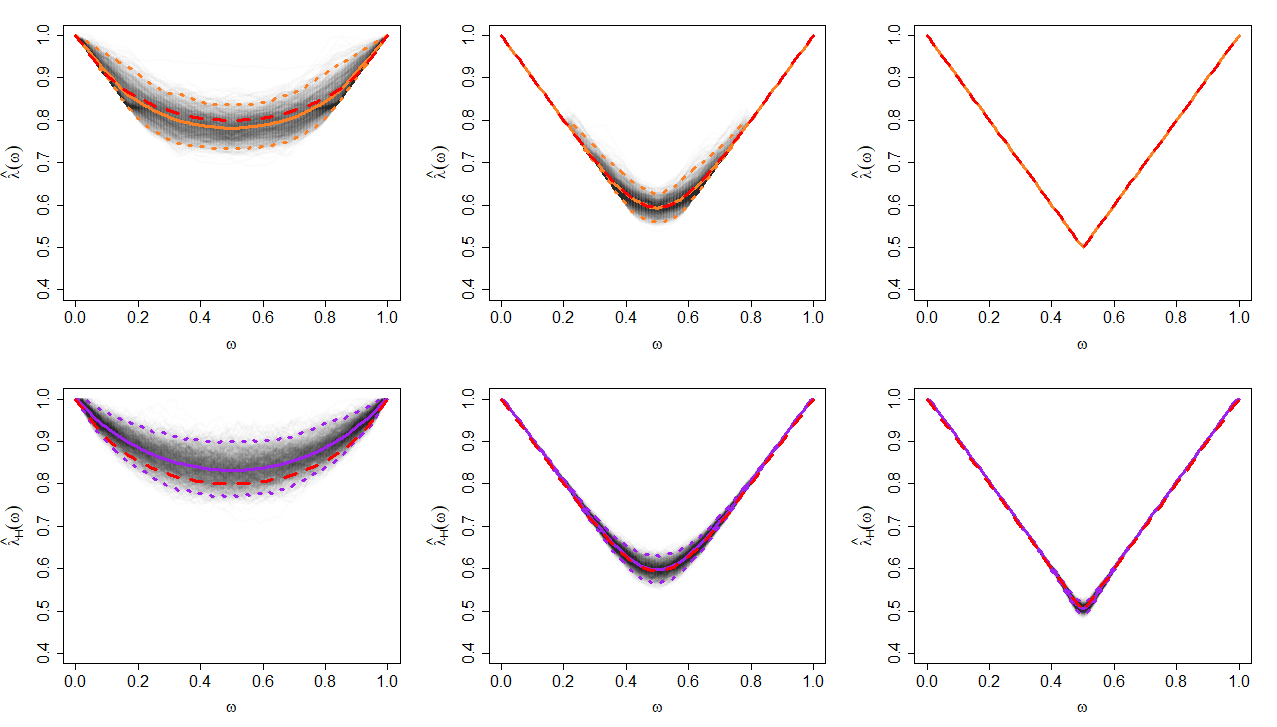}
        \caption{Estimates of $\lambda(\omega)$, $\omega\in[0,1]$, for data simulated from Gaussian (left), inverted logistic (centre) and logistic (right) models with their corresponding parameters set as 0.25. Each grey line represents an estimate obtained using $\hat\lambda_G(\omega)$ (top) or $\hat\lambda_{H}(\omega)$ (bottom) over 1000 simulations. The solid and lower/upper dotted lines show the pointwise bootstrapped means and 0.025 and 0.975 quantiles, respectively, for $\hat\lambda_G(\omega)$ (orange) and $\hat\lambda_H(\omega)$ (purple). The true $\lambda(\omega)$ values are shown in red.}
    \label{fig:lambdaEstimates0.25} 
\end{figure}

\begin{figure}[!htbp]
    \centering
    \includegraphics[width=0.95\textwidth]{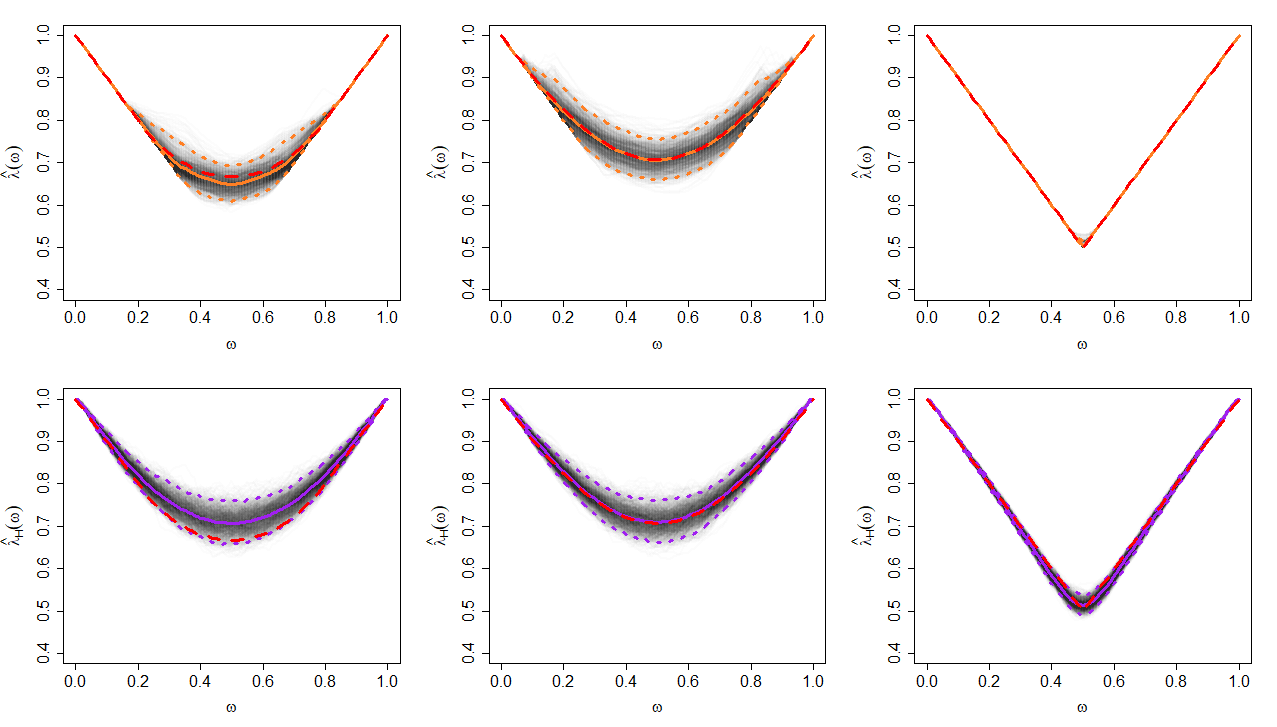}
        \caption{Plots equivalent to Figure~\ref{fig:lambdaEstimates0.25}, but with the copula parameters set to 0.5.}
    \label{fig:lambdaEstimates0.50}    
\end{figure}

\begin{figure}[!htbp]
    \centering    
    \includegraphics[width=0.95\textwidth]{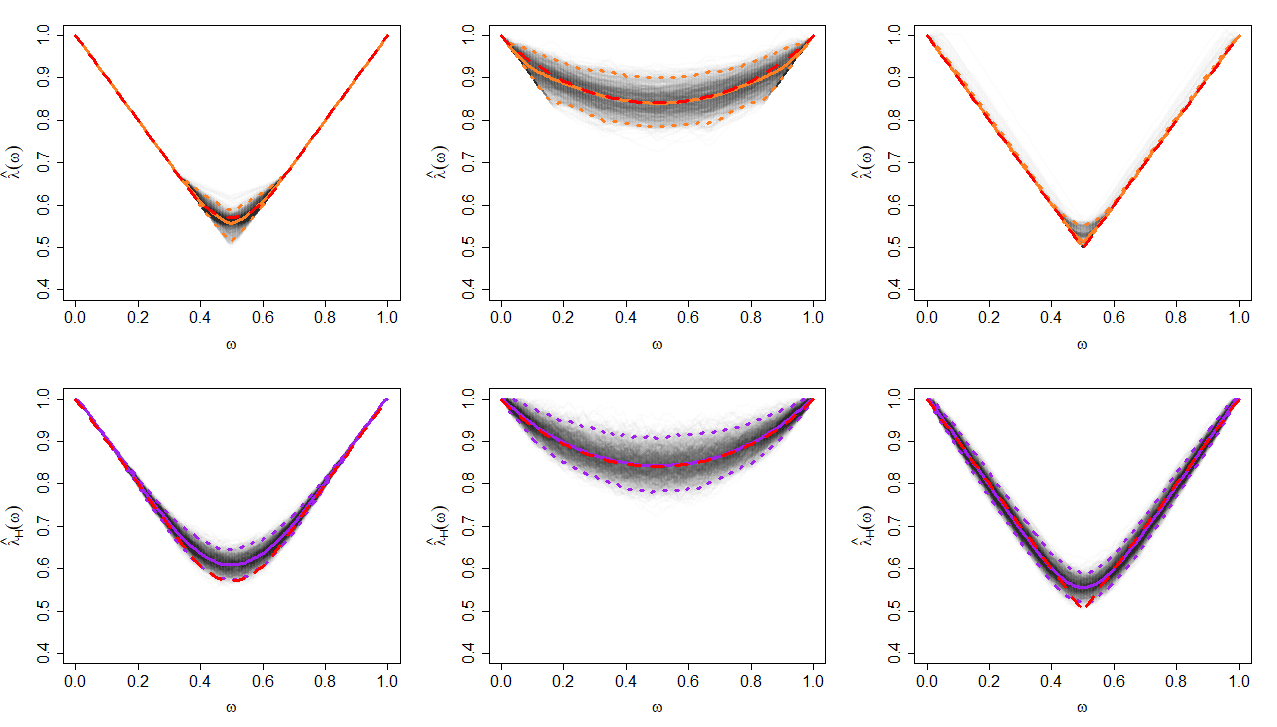}
        \caption{Plots equivalent to Figure~\ref{fig:lambdaEstimates0.25}, but with the copula parameters set to 0.75.}
    \label{fig:lambdaEstimates0.75}
\end{figure}

We now present results on the estimation of $\lambda(\omega)$. The simulation results presented in Figures~\ref{fig:lambdaEstimates0.25},~\ref{fig:lambdaEstimates0.50}~and~\ref{fig:lambdaEstimates0.75} are analogous to those presented in Figures~\ref{fig:tauEstimates0.25},~\ref{fig:tauEstimates0.5} (in the main paper)~and~\ref{fig:tauEstimates0.75} in the $\tau_1(\delta)$ case, respectively, i.e., for Gaussian, inverted logistic and logistic data with $\rho,\gamma\in\{0.25,0.5,0.75\}$. For the estimator $\hat\lambda_H(\omega)$, we set the threshold $u_\omega$ as the observed 0.95 quantile of the structure variable $M_\omega$ in each case. For the Gaussian examples, $\hat\lambda_H(\omega)$ has a tendency towards overestimation; this bias is significantly reduced by our proposed estimator. In the case of the inverted logistic data, the two approaches give comparable, unbiased results with similar variability. Finally, for the logistic examples, with $\gamma=0.25,0.50$, our approach is particularly successful, giving almost perfect estimates of $\lambda(\omega)$ across the full range of $\omega\in[0,1]$. For the logistic model with $\gamma=0.75$, we also obtain strong results that are an improvement over the Hill-type estimator.

\section{Estimates of $(\alpha_1,\beta_1)$}\label{SM:HTResults}
\begin{figure}[!htbp]
    \centering
    \includegraphics[width=0.75\textwidth]{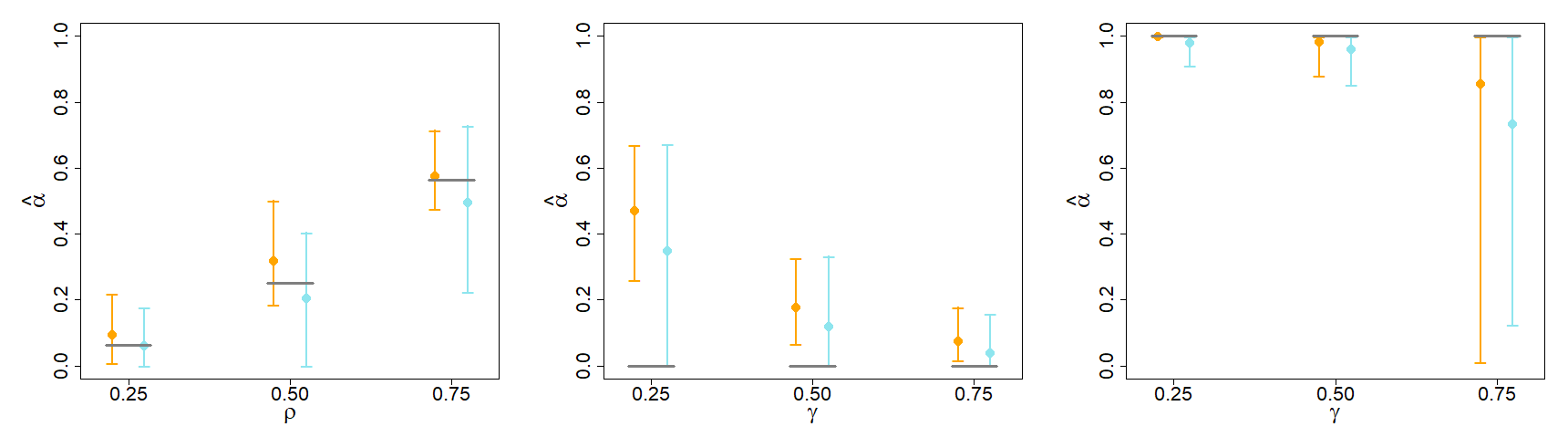}
    \includegraphics[width=0.75\textwidth]{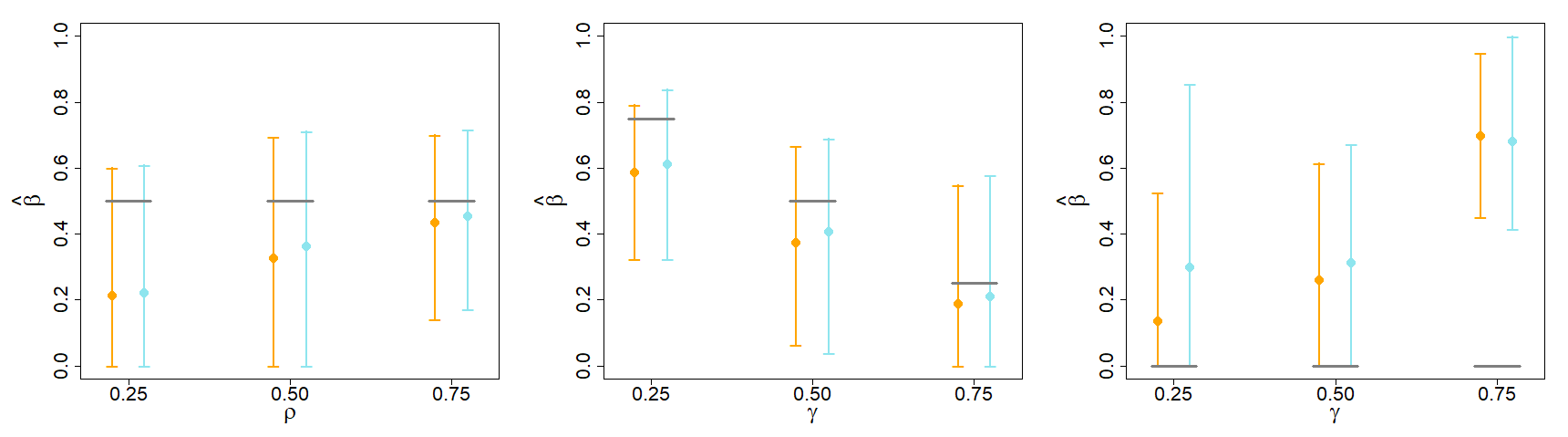}
    \caption{Estimates of $\alpha_1$ (top row) $\beta_1$ (bottom row) for data simulated from Gaussian (left), inverted logistic (centre) and logistic (right) models with their corresponding parameters taken as 0.25, 0.5 and 0.75. We show the mean estimates of $\beta_1$ (circles) and 0.025 to 0.975 quantile range over 1000 simulations for $\hat\beta_{G,1}$ (orange) and maximum likelihood estimates of $\beta_1$. The true $\beta_1$ values are shown in grey.}
    \label{fig:betaEstimates}
\end{figure}

In Figure~\ref{fig:betaEstimates}, we present results on the estimation of $\alpha_1$ and $\beta_1$ from the conditional extremes model of \cite{Heffernan2004}. These are compared to results obtained using maximum likelihood estimators for model~\eqref{eqn:HTestimate} for both parameters simultaneously. Our results are reasonably similar to those obtained using the maximum likelihood estimators, but slightly worse for inverted logistic data, although we do offer some improvement in the logistic case with $\gamma=0.25,0.5$. Most noticeably, neither approach is able to provide very successful estimation of the $\beta_1$ parameter, but this is known to be difficult to estimate.

\section{Asymmetric logistic example}\label{SM:alog}
The three copulas we considered in Section~\ref{subsec:simulationStudy} of the main paper exhibit either asymptotic independence, or asymptotic dependence where both variables can only take their largest values simultaneously. We now demonstrate our approach for the asymmetric logistic model \citep{Tawn1988}, which can exhibit a mixture structure in its extremal dependence features, as discussed in Section~\ref{sec:NWtheory}. 

We first provide more detail on the form of this model. In exponential margins, the bivariate extreme value distribution function has the form
\[
F(x_1,x_2)=\exp\left[-V\left\{\frac{-1}{\log(1-e^{-x_1})},\frac{-1}{\log(1-e^{-x_2})}\right\}\right]
\]
for $x_1,x_2\geq 0$ and an exponent measure $V(x,y)$ satisfying certain conditions. For the asymmetric logistic model, 
\[
V(x,y) = \frac{\theta_1}{x} + \frac{\theta_2}{y} + \left\{\left(\frac{1-\theta_1}{x}\right)^{1/\gamma}+\left(\frac{1-\theta_2}{y}\right)^{1/\gamma}\right\}^\gamma,
\]
with $x,y>0$, $\gamma\in(0,1]$ and $\theta_1,\theta_2\in[0,1]$. For this model, we have 
\[
\chi = 2-\theta_1-\theta_2-\left\{(1-\theta_1)^{1/\gamma}+(1-\theta_2)^{1/\gamma}\right\}^\gamma.
\]
For a fixed $\gamma<1$, the parameters $(\theta_1,\theta_2)$ control the extremal dependence structure of the variables $(X_1,X_2)$ and the strength of asymptotic dependence. If $\theta_1>0$, it is possible for $X_1$ to be large while $X_2$ is small, and an analogous result holds for $\theta_2>0$. If $\max(\theta_1,\theta_2)<1$, then both variables can take their largest values simultaneously. In the case where $\theta_1,\theta_2\in(0,1)$, the asymmetric logistic model can be thought of as a mixture of a logistic model with independence. Asymptotically, proportions $\theta_1/2,\theta_2/2$ and $1-(\theta_1+\theta_2)/2$ of the extremal mass are associated with the cases where $X_1$ is large while $X_2$ is small, $X_2$ is large while $X_1$ is small, and $(X_1,X_2)$ are simultaneously large, respectively. The gauge function of this model does not depend on $(\theta_1,\theta_2)$, and takes the form~\eqref{eqn:alogGauge} in the main paper.

\begin{figure}[t]
    \centering
    \includegraphics[width=0.95\textwidth]{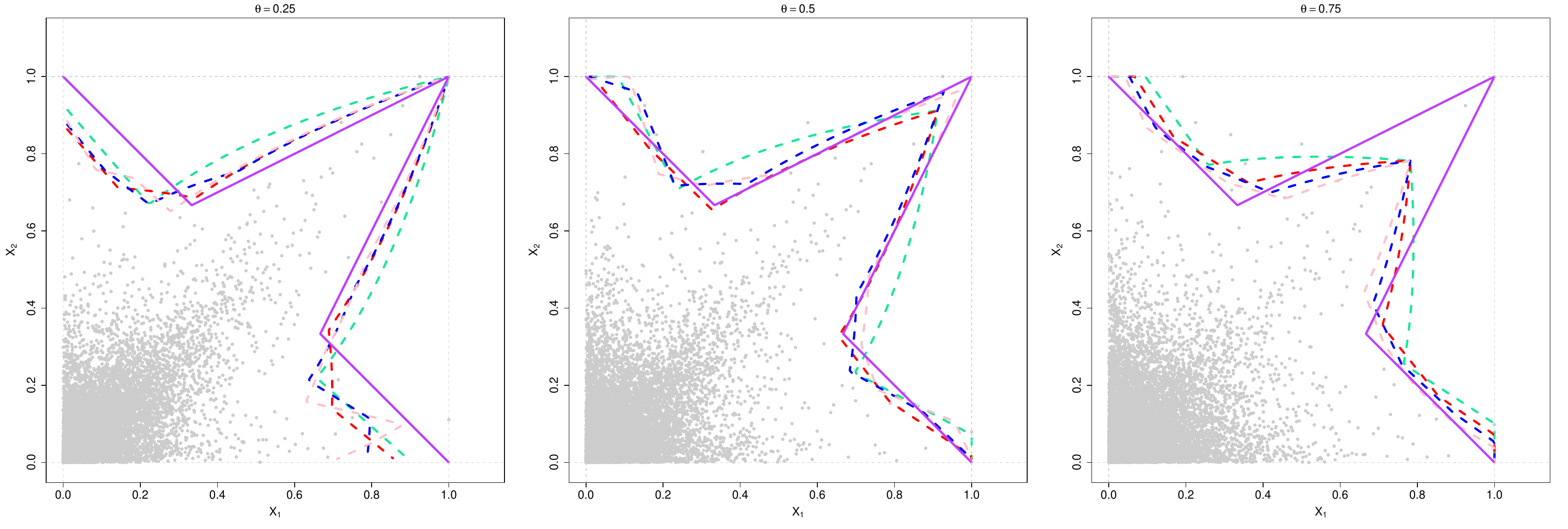}
    \caption{An example of our estimation procedure for the asymmetric logistic model with $\gamma=0.5$ and $\theta\in\{0.25,0.5,0.75\}$. The true set $G$ is demonstrated by the purple lines; the simulated data used to obtain the estimates of $G$ are shown in grey; the dashed lines show our estimates of $G$ for different numbers of knots in the spline functions ($\kappa=5$: green; $\kappa=7$: red; $\kappa=9$: blue; $\kappa=11$: pink).}
    \label{fig:alogExamples}
\end{figure}

We apply our estimation procedure to three samples of size 10,000 from the asymmetric logistic model with different values of $\theta_1$ and $\theta_2$ (which we take to be equal). If $\theta_1=\theta_2=\theta$, we have $\chi=(2-2^\gamma)(1-\theta)$, i.e., the strength of asymptotic dependence decreases as the value of $\theta$ increases. We fix the tuning parameters as in Section~\ref{subsec:simulationStudy}, and the resulting $\hat G$ estimates are shown by the red dashed lines in Figure~\ref{fig:alogExamples}. This estimated boundary set is reasonably close to the truth for $\theta=0.25$ and $\theta=0.5$, where a respective 75\% and 50\% of the extremal mass is associated with both variables being simultaneously large \citep{Simpson2020}. When $\theta=0.75$, so the proportion of mass associated with the simultaneously extreme variables is reduced to 25\%, it becomes more difficult for $\hat G$ to detect this feature. 

We also carry out our estimation procedure using different numbers of knots $\kappa$ in the spline functions of the GAMs, as here $G$ is not smooth everywhere and increasing the number of knots may enable us to better capture this feature. Results with $\kappa\in\{5,9,11\}$ are shown in Figure~\ref{fig:alogExamples}. The results for $\kappa=5$ are indeed slightly worse than for $\kappa\in\{7,9,11\}$, but otherwise our estimates of $G$ are not particularly sensitive to this tuning parameter. An issue highlighted in Figure~\ref{fig:alogExamples} is that our approach to scaling the boundary estimate onto $[0,1]^2$ means we are unlikely to have our estimate of $G$ intersecting each of the lines $x_1=1$ and $x_2=1$ in more than one region. As a result, it is difficult for our approach to detect extremal dependence mixture structures. We expand further on this point in our discussion in Section~\ref{subsec:issues} of the main paper.

\begin{figure}[!htbp]
        \centering
        \includegraphics[width=0.95\textwidth]{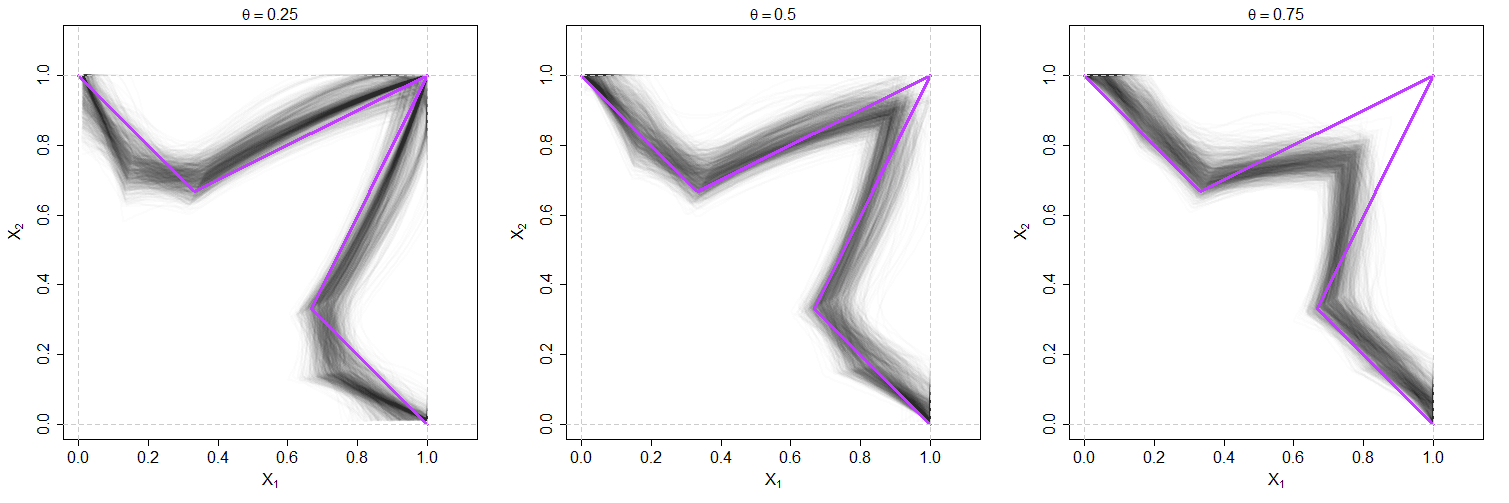}
        \caption{Estimates $\hat G$ for the asymmetric logistic copula with $\gamma=0.5$ and $\theta\in\{0.25,0.5,0.75\}$ (left to right). The sample size in each case is $n=10,000$ and 1000 estimates are shown for each model (grey). The true sets $G$ are shown in purple.}
        \label{fig:alogExtra}
    \end{figure}

In Figure~\ref{fig:alogExtra}, we present additional results over 1000 repetitions for the default choice of $\kappa=7$, with the remainder of the tuning parameters fixed as in Section~\ref{subsec:simulationStudy}. The estimates are similar to those given in Figure~\ref{fig:alogExamples}, with a similar amount of variability as in the Gaussian, inverted logistic and logistic cases of Section~\ref{subsec:simulationStudy}. In particular, we see that when $\theta=0.75$, it remains difficult to completely capture the larger mixture component when $X_1$ is large for each sample replicate.

\section{Results for a larger sample size}\label{SM:size100000}
\begin{figure}[!htbp]
    \centering
    \includegraphics[width=0.95\textwidth]{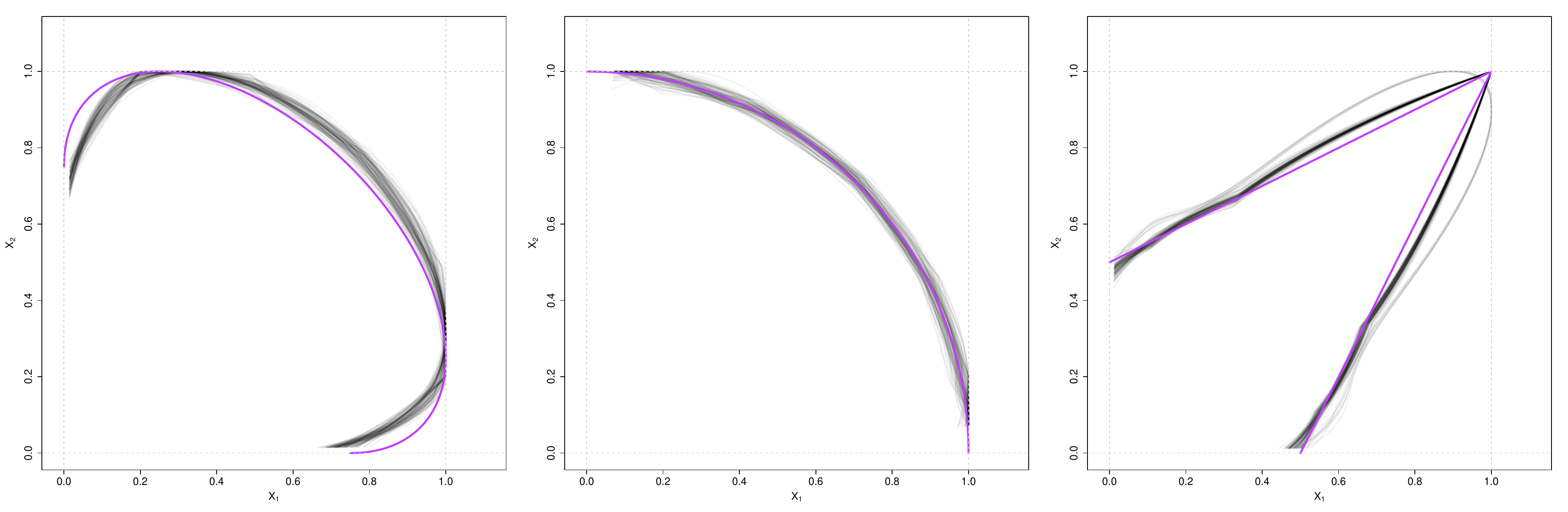}
    \caption{Estimates $\hat G$ for the Gaussian (left), inverted logistic (centre) and logistic (right) models, with the corresponding $\rho$ or $\gamma$ parameters set to 0.5. The sample size in each case is $n=100,000$ and 100 estimates are shown for each model (grey). The true sets $G$ are shown in purple.}
    \label{fig:Gexample_n100000}
\end{figure}

In the simulation studies of the main paper, and thus far in this Supplementary Material, we have considered only samples of size $n=10,000$, both for consistency and to ensure the sample sizes were of a similar order to that of the wave height application in Section~\ref{subsec:waves} of the main paper. We now provide results for simulated datasets of size $n=100,000$ to check the reliability of the procedure for different sample sizes. Estimates of the set $G$ over 100 iterations for the three copula models studied in Section~\ref{subsec:simulationStudy} of the main paper are shown in Figure~\ref{fig:Gexample_n100000}; all tuning parameters are fixed as in Section~\ref{subsec:simulationStudy}. The spline degrees chosen by our algorithm are summarised in Table~\ref{tab:splineResultsn100000}.

The results in Figure~\ref{fig:Gexample_n100000} and Table~\ref{tab:splineResultsn100000} show that the tuning parameters we have selected give reasonable results when the sample size is increased to $100,000$. For all three copula types, the variance of our estimates has decreased compared to the results for samples of size $10,000$. When considering the degree of the splines chosen in the GAM component of the method, it is noticeable that the most appropriate linear splines are chosen more often for the logistic model when the sample size is increased. It is also reassuring that non-linear options are chosen more often for the Gaussian model as the sample size is increased, although the effect of sample size on the choice of spline degree appears negligible in the inverted logistic case.

\begin{table}[!htbp]
    \centering
    \begin{tabular}{c|ccc}
        Spline degree &  Gaussian  &  Inverted logistic  &  Logistic \\
        \hline 
        1 &  22  &  36  &  91  \\
        2 &  41  &  42  &   0  \\
        3 &  37  &  22  &   9  \\
    \end{tabular}
    \caption{The number of times each spline degree is chosen for the estimated sets $\hat G$ shown in Figure~\ref{fig:Gexample_n100000}.}
    \label{tab:splineResultsn100000}
\end{table}

\begin{figure}[!htbp]
    \centering
    \includegraphics[width=0.95\textwidth]{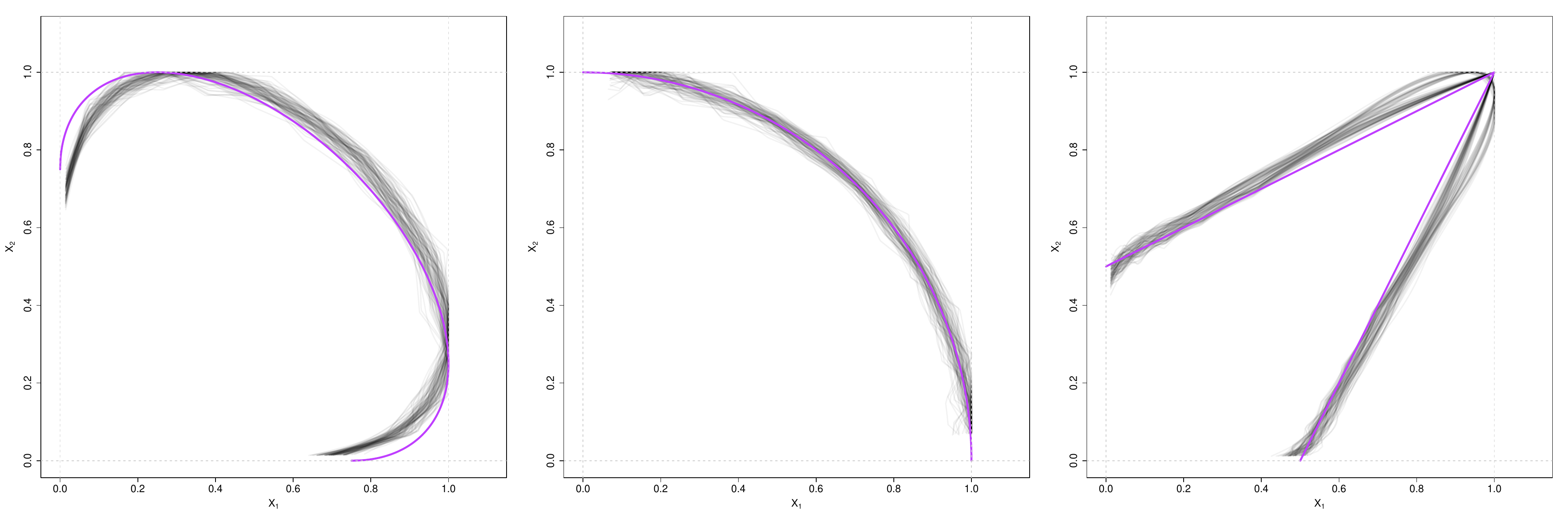}
    \caption{Estimates $\hat G$ with $\kappa=15$ for the Gaussian (left), inverted logistic (centre) and logistic (right) models, with the corresponding $\rho$ or $\gamma$ parameters set to 0.5. The sample size in each case is $n=100,000$ and 100 estimates are shown for each model (grey). The true sets $G$ are shown in purple.}
    \label{fig:Gexample_n100000_k15}
\end{figure}

There remains some bias in the Gaussian examples in Figure~\ref{fig:Gexample_n100000}. This raises the question of whether additional spline knots are required as the sample size increases. In Figure~\ref{fig:Gexample_n100000_k15}, we provide further estimates of the set $G$ with the number of knots increased to $\kappa=15$. Increasing the number of knots has not improved the Gaussian results, but has led to a deterioration of the estimates in the logistic case, where linear splines are now chosen in only 60\% of cases. This suggests having a smaller number of knots, as suggested in the main paper, is reasonable across the range of copula examples and sample sizes we consider.

The bias in the Gaussian example may be due to slow convergence to the limit set for this class of copula. This conclusion is consistent with existing theoretical and empirical studies of convergence in extremal properties for the Gaussian copula. For example, \cite{Bofinger1965} show that convergence of bivariate maxima for the Gaussian copula is very slow; for the coefficient of asymptotic independence, $\eta$ \citep{Ledford1996,Ledford1997}, the penultimate extremal index is slow to converge to its limit value of $1$ in the Gaussian case \citep{Ledford2003}; and there is slow convergence for the conditional extremes distribution \citep{Heffernan2004,Lugrin2021}. In all cases the convergence rates are controlled by slowly varying functions, which are $O\left[(\log u)^{-\rho/(1+\rho)}\right]$, where $\rho$ is the correlation parameter and $u$ is the threshold. This bias is also evident in our application of existing estimators of the extremal dependence properties to the Gaussian copula in the main paper. Given that each of these extremal characteristics is determined by the set $G$, it is not at all surprising that the convergence in this case is very slow. This is not to say that the bias in estimating $G$ is too problematic for statistical inference/extrapolation purposes, as at any typical level of extrapolation the bias will persist and the resulting inference will provide reliable estimates for finite quantities of interest.

\section{Coverage analysis for the non-parametric bootstrap}\label{SM:bootstrap}
\begin{figure}[!htbp]
    \centering
    \includegraphics[width=0.55\textwidth]{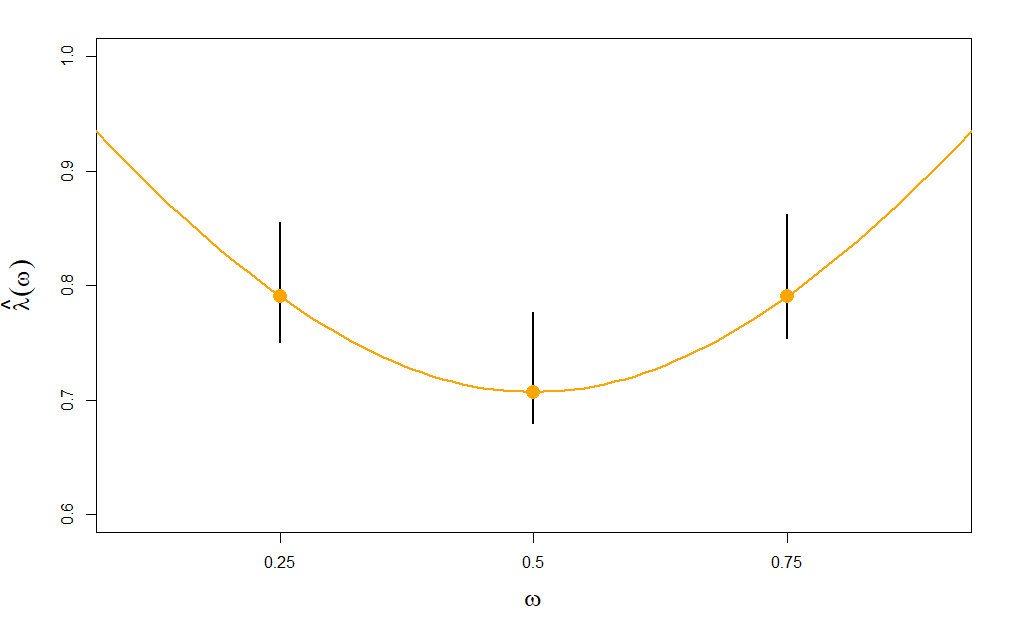}
    \caption{An example of the 95\% bootstrapped confidence intervals for $\lambda(0.25)$, $\lambda(0.5)$ and $\lambda(0.75)$ (black lines) for a sample of size $n=10,000$ taken from an inverted logistic model with $\gamma=0.5$. The true $\lambda(\omega)$ values for this model are shown in orange.}
    \label{fig:bootstrap}
\end{figure}

In this section, we demonstrate the performance of the non-parametric bootstrap for the assessing uncertainty in our estimates. We focus on one copula, namely the inverted logistic model with $\gamma=0.5$. All tuning parameters are fixed as in Section~\ref{subsec:simulationStudy} of the main paper. We focus on the coverage of 95\% confidence intervals with respect to the dependence measure $\lambda(\omega)$ for $\omega=0.25,0.5,0.75$. The set up of our study is as follows:
\begin{enumerate}
    \item We simulate a dataset of size $n=10,000$ from our inverted logistic model, then estimate $G$ and extract the corresponding estimates of $\lambda(\omega)$ for $\omega=0.25,0.5,0.75$.
    \item We generate 250 bootstrapped samples of size $10,000$, by sampling with replacement from the dataset in step 1. For each of these samples, we again estimate $\lambda(\omega)$ for $\omega=0.25,0.5,0.75$.
    \item We extract the 0.025 and 0.975 percentiles from our bootstrapped estimates of $\lambda(\omega)$, for each $\omega=0.25,0.5,0.75$, providing a 95\% bootstrapped confidence interval in each case.
    \item We repeat steps 1-3 for 300 times, so we have 300 95\% bootstrapped confidence intervals for each $\lambda(\omega)$.
\end{enumerate}
In Figure~\ref{fig:bootstrap}, we show an example of the 95\% confidence intervals obtained for one sample from our inverted logistic model (i.e., after carrying out steps 1-3 above just once). In this case, all three confidence intervals for $\lambda(\omega)$ $(\omega=0.25,0.5,0.75)$ contain the truth. Considering the results across all 300 samples, 99\% of the $\lambda(0.25)$ confidence intervals, 92\% of the $\lambda(0.5)$ confidence intervals and 99\% of the $\lambda(0.75)$ confidence intervals contain the truth, suggesting that we have reasonable coverage with respect to the target of 95\%.

\section{The algorithm for estimating $G$}\label{SM:algorithm}
In Section~\ref{sec:method} of the main paper, we present a detailed discussion of our approach for estimating the set $G$. Here, we provide a summarised version which may be more convenient for the reader wishing to implement our procedure. Our \texttt{R} code can also be found online in the GitHub repository \url{https://github.com/essimpson/self-consistent-inference}. The approach is as follows:

\begin{enumerate}
    \item Apply the rank transformation to obtain exponential margins as in equation~\eqref{eqn:rankTransform}.
    \item Transform the data to pseudo-polar coordinates as in~\eqref{eqn:pseudo-polar}, with observations denoted $(r_1,w_1),\dots,(r_n,w_n)$.
    \item Select the angles $w^*_j$, $j\in J_k:=\{1,\dots,k\}$, at which to estimate high radial quantiles. 
        \begin{itemize}
            \item {\bf Default tuning parameter:} $k=199$.
            \item {\bf Default angle selection:} set $w_j^*$ as the $(j-1)/(k-1)$th empirical quantile of $\{w_1,\dots,w_n\}$, for $j=1,\dots,k-1$, and $w_k^*=1/2$. (Note that this labelling leads to unordered $w_j^*$ values.)
        \end{itemize}
    \item For each $w_j^*$: 
       \begin{enumerate}
            \item Select $\epsilon_{w_j^*}$ in~\eqref{eqn:Rneighbours} such that $|\mathcal{R}_{w_j^*}|=m$.
                \begin{itemize}
                    \item {\bf Default tuning parameter:} $m=100$.
                \end{itemize}
            \item Select the GPD threshold $u_{w_j^*}$ in~\eqref{eqn:GPD} as the empirical $q_u$th quantile of the values in $\mathcal{R}_{w_j^*}$. Estimate the parameters $\sigma(w_j^*)$ and $\xi(w_j^*)$ in~\eqref{eqn:GPD} by applying maximum likelihood estimation to the observed radial values in $\mathcal{R}_{w_j^*}$ exceeding $u_{w_j^*}$.
                \begin{itemize}
                    \item {\bf Default tuning parameter:} $q_u=0.5$.
                \end{itemize}
            \item Estimate the high radial quantile $r^L_q(w^*_j)$ via equation~\eqref{eqn:radialQuantile}.
                \begin{itemize}
                    \item {\bf Default tuning parameter:} $q=0.999$.
                \end{itemize}
        \end{enumerate}
    \item Choose the position of the $\kappa$ spline knots to be used in the smooth GPD parameter estimates.
        \begin{itemize}
            \item {\bf Default tuning parameter:} $\kappa=7$.
            \item {\bf Default knot selection:} take knots evenly spaced within the range of observed angles, but with the central one adjusted to be exactly $1/2$.
        \end{itemize}
    \item For each $d\in\{1,2,3\}$:
        \begin{enumerate}
            \item Use an asymmetric Laplace distribution with a spline of degree $d$ to estimate smooth GPD thresholds $u_w$ at the quantile level $q_u$ (selected above).
                \begin{itemize}
                    \item {\bf Note:} this distribution is fitted to the log-radial values, with the required quantile extracted and back-transformed onto the radial scale.
                \end{itemize}
            \item Fit a GPD-GAM model to the radial threshold excesses, using a spline of degree $d$ for the log-scale parameter.
                \begin{itemize}
                    \item {\bf Default:} constant shape parameter. 
                \end{itemize}
            \item Estimate the high radial quantiles $r^{S_{d^*}}_q(w^*_j)$ $(j\in J_k)$ (with $q$ selected as above).
            \item Obtain an estimate $\hat G^{S_d}$ of $G$ by scaling onto $[0,1]^2$ using the procedure outlined in Section~\ref{SM:estimationIllustration}.
        \end{enumerate}
    \item Choose $G$ by minimising the mean absolute difference of the radial quantile estimates as in~\eqref{eqn:Gest}.
\end{enumerate}

\end{document}